\documentclass[11pt,arial,a4paper]{article}
\usepackage[utf8]{inputenc}
\usepackage[T1]{fontenc}
\usepackage{lmodern}
\usepackage{alphabeta}
\usepackage{authblk}
\usepackage{graphicx}
\usepackage{amssymb}
\usepackage{float}
\usepackage[caption = false,position=top]{subfig}
\usepackage{setspace}
\usepackage{multirow}
\usepackage{amsmath}
\usepackage{url}
\usepackage[
  backend=biber,
  style=apa,
  sorting=nyt,
  natbib=true
]{biblatex}
\usepackage{color,soul}
\usepackage[margin=.75in]{geometry}
\usepackage{bm}
\usepackage{dsfont}
\usepackage[svgnames]{xcolor}

\newcommand*\vect[1]{\mathbf{\bm{#1}}}

\usepackage{soul}

\newcommand{\y}{\vect{y}}
\newcommand{\X}{\vect{X}}
\newcommand{\Xo}{\X_0}
\newcommand{\Xl}{\X_1}
\newcommand{\al}{\vect{\alpha}}
\newcommand{\be}{\vect{\beta}}

\begin{document}

\title{Bayesian Evidence Synthesis for the common effect model  \footnote{\emph{Acknowledgements}: This work was supported by the DRASI-2 funding scheme, Athens University of Economics and Business}}

\author[1,2]{Stavros Nikolakopoulos\footnote{corresponding author; e:snikolakopoulos@uoi.gr}}
\author[3]{Björn Alfons Edmar}
\author[4]{Ioannis Ntzoufras}
\affil[1]{Department of Psychology, University of Ioannina}
\affil[2]{Department of Biostatistics, University Medical Center Utrecht}
\affil[3]{Department of Statistics, Athens University of Economics and Business}
\affil[4]{Department of Methodology and Statistics, Utrecht Unversity}

\date{}
\maketitle

\begin{abstract}
Bayes Factors, the Bayesian tool for hypothesis testing, are receiving increasing attention in the literature. Compared to their frequentist rivals ($p$-values or test statistics), Bayes Factors have the conceptual advantage of providing evidence both for and against a null hypothesis, and they can be calibrated so that they do not depend so heavily on the sample size. Recently, research on the synthesis of Bayes Factors arising from individual studies has received increasing attention, mostly for the fixed effects model for meta-analysis. In this work, we review and propose methods for combining Bayes Factors from multiple studies, depending on the level of information available, focusing on the common effect model. In the process, we provide insights with respect to the interplay between frequentist and Bayesian evidence. We assess the performance of the methods discussed via a simulation study and apply the methods in an example from the field of positive psychology. 

\end{abstract}
\pagebreak
\singlespacing
\pagestyle{myheadings}
\markboth{}{Bayesian Evidence Synthesis for the common effect model}

\newpage

\begin{center}
\Large{\textbf{Bayesian Evidence Synthesis for the common effect model}}
\end{center}

\section{Introduction} 
Evidence synthesis is important for generalizing, summarizing, and evaluating empirical findings. The choice of synthesis method relies on many factors, such as the overarching research question, the format and availability of data, and the expertise and methodological conventions in the given field. Evidence synthesis - like most scientific methodology - can be approached using both qualitative and quantitative methods. 

Qualitative evidence synthesis, typically in the form of a review (\cite{carrera-rivera_how-conduct_2022}), collects all relevant studies for a given research question and summarizes the findings descriptively. A purely qualitative approach to evidence synthesis is taken when studies lack comparability (e.g., few replications, no shared paradigm) or when essential statistical information is missing from the reporting. Although qualitative synthesis is not inferior to quantitative analysis, especially in the social sciences ‐, where abstract constructs such as cognition, emotion, attitudes, and behavior are of concern; its reliance on narrative description restricts the ability to quantify the cumulative strength of evidence, formally compare competing hypotheses, or assess heterogeneity and publication bias.

In contrast, quantitative evidence synthesis, provides a principled framework for cumulating evidence across studies, yielding explicit measures of support for or against theoretical claims, and enabling a more transparent, reproducible, and cumulative science. It stands on the shoulders of the qualitative review, but where enough information and similarity across studies has been established to allow for a statistical model to be fit to the gathered data (studies). Quantitative synthesis is generally called a "meta-analysis". Importantly, meta-analysis is not a single modeling technique or approach, but rather a family of statistical models that differ in how they conceptualize the relationship between observed and true effects. These models are commonly referred to as the common-effect model, the random-effects model, and the fixed-effects model (\cite{mckenzie_brief_2024,rice_re-evaluation_2018}). These models form the statistical foundation for  parameter estimation and hypothesis testing in evidence synthesis. In the following section we will detail these models more closely by describing the assumptions they make and the hypotheses they test. 

The naming convention of these three models have been used inconsistently across disciplines, thus we aim to be as explicit as possible about our terminology and therefore clarify our usage here. The common-effect model assumes a single shared true effect size across all studies: $y_i \sim \mathcal{N}(\theta, \sigma^2_i)$. The random-effects model assumes each study's effect come from a distribution of true effects: $y_i \sim \mathcal{N}(\theta_i, \sigma^2_i)$, where $\theta_i \sim \mathcal{N}(\mu, \tau^2)$. The fixed-effects model assumes that each study has its own true effect $\theta_i$, which is considered fixed and not random: $y_i \sim \mathcal{N}(\theta_i, \sigma^2_i)$. In the realm of general linear models, this can be expressed by a study*treatment interaction. Clearly, different modeling assumptions result to different methods needed for the estimation of the relevant treatment effect(s).

The common- and random-effects meta-analytic models are the most common quantitative synthesis method in the social sciences. They are appropriate when substantial similarity exists across a body of literature and the research question focuses on estimating a parameter, such as a standardized mean difference or correlation/regression coefficient. Using this approach, effect sizes and standard errors are extracted from individual studies and synthesized using meta-analytic pooling. Despite the variety of available models , they all share the goal of estimating a typical effect within a population of studies, where the included studies are assumed to represent a random sample from this broader population (\cite{zhang_towards_2023}).

 It is important to note that while the different meta-analytic models are similar in statistical formulation, they test different hypotheses and estimate different parameters. This is a common source of confusion when distinguishing between the common-effect model and fixed-effects model. When the objective is to estimate an overall effect size, the point estimate for the common-effect model and the fixed-effects model will be identical (\cite{hedges_fixed-_1998},\cite{rice_re-evaluation_2018}, \citet{veroniki_brief_2024}). This is because the estimation formula is the same, although the interpretation of the estimate differ. Since the common-effect model assumes that all observed effects come from the same distribution $y_i \sim \mathcal{N}(\theta, \sigma^2_i)$, it is estimating the unconditional true effect in a homogeneous study population. The fixed-effects model makes no assumption regarding the homogeneity of the effects, and therefore simply estimates the weighted average true effect of the studies included in the analysis. In essence, this difference boils down to different assumptions about heterogeneity in effects and whether you make unconditional (common-effect) or conditional (fixed-effects) inferences regarding the effect size. 

When the focus is on evaluating hypotheses concerning effects, the different formalization results in the assessment of different hypotheses. Through a frequentist null hypothesis testing perspective, the common-effect model evaluates the null hypothesis that the shared true effect in the population is zero (i.e $\theta = 0$). In the random-effects model the null hypothesis is that the average of all individual true effects is zero (i.e $\mu = 0$), and in the fixed-effects model the null hypothesis is that all individual true effects are zero (i.e $\theta_i = 0, \forall i$).

Recently, there has been renewed interest in using Bayes factors (BF) to synthesize evidence specifically for the fixed-effects model, assessing whether a given hypothesis holds consistently across studies (\cite{klugkist_bayesian_2023}, \cite{van_lissa_tutorial_2024}, \cite{van_wonderen_bayesian_2024}). This shift from frequentist null hypothesis testing to Bayesian model comparison using Bayes Factors clarifies the distinction between the common-effect and fixed-effects models (\cite{mulder_bayesian_2024}). The common-effect model assessing whether the shared parameter is the same across all studies, and the fixed-effects model assessing whether all studies share the same hypothesis structure or direction.

In this paper we expand on Bayesian evidence synthesis using the common-effects model and provide a new method to calculate a pooled BF across a set of available studies. We focus on the simple linear regression model with either a continuous or a binary covariate, the latter representing the very common scenario of synthesizing t-tests from randomized studies. Like the approach by \citet{rouder2011}, we derive an analytic expression connecting Bayes Factors with frequentist test statistics, integrated within a meta-analytic framework. We also apply the methods in a published meta-analysis. 

\section{Preliminaries}
\subsection{The Bayes Factor}
For  a given set of response values $\y$ and two competing models $M_0=\{f(\cdot|\theta_0,M_0),\theta_0\in \Theta_0\}$ and $M_1=\{f(\cdot|\theta_1,M_1),\theta_1\in \Theta_1\}$ the Bayes Factor $B_{10}$ is the ratio of the marginal densities of $\y$ under $M_1$ and $M_0$ given by 
\begin{align*}
BF_{10}={\frac {f(\y|M_{1})}{f(\y|M_{0})}}={\frac {\int_{\Theta_1} f(\y|\theta_{1},M_{1})\pi(\theta_{1}|M_{1})d\theta_{1}}{\int_{\Theta_0} f(\y|\theta_{0},M_{0})\pi(\theta_{0}|M_{0})\,d\theta _{0}}}
\end{align*}
where $\pi(\theta_\ell|M_\ell)$ represents the prior distribution of $\theta_\ell$ under model $M_\ell$, for $\ell \in \{0,1\}$. Note also that
\begin{align*}
PO_{10} = \frac{Pr(M_1|\y)}{Pr(M_0|\y)}=BF_{10}\times \frac{Pr(M_1)}{Pr(M_0)}
\end{align*}
where $PO_{10}$ denote the posterior model odds for $M_1$ versus $M_0$, so $BF_{10}$ quantifies the increase in the odds of $M_1$ relative to $M_0$ after the data is observed.

The Bayes factor is a special case of posterior model odds when considering equal prior probabilities for each hypothesis or model under consideration. 
Even though $PO$ is a natural tool for comparing two hypotheses or models in the Bayesian framework \citep{jeffreys1961,kass1995}, here we focus on the synthesis of $BF$s rather than $PO$s for several reasons. 
First, the choice of equal probabilities for the two competing models/hypotheses is considered as a reasonable non-informative choice.
Moreover, when synthesizing Bayesian evidence, it is sensible to leave any study specific prior preferences outside of the meta-analysis procedure. Thus, we estimate a combined $BF$ across all studies and then incorporate any overall prior model preferences at the end of the meta analytic procedure. 

Since the $BF$ does not depend on the prior probabilities of the models, it is usually interpreted as the odds provided by the data for $M_1$ to $M_0$. However, $BF_{10}$ obviously depends on the prior distributions of the model parameters. Large values of $BF_{k\ell}$ indicate strong posterior support of model $M_k$ against model $M_\ell$; for details see, e.g. \citet{kass1995}.

\subsection{Bayes Factors with conjugate priors}\label{sec:2.2}
We consider the standard general hypothesis testing setting in linear regression with normal independent errors. 

\begin{equation}\label{eq:2.1}
\begin{aligned}
H_0: \be=0     \Rightarrow M_0 &: \y| \al,     \tau \sim N_n\left( \Xo \al ,\tau^{-1}\vect{I}_n\right) \\
H_1: \be\neq 0 \Rightarrow M_1 &: \y| \al, \be,\tau \sim N_n\left( \Xo \al +\Xl \be ,\tau^{-1}\vect{I}_n\right) 
\end{aligned}
\end{equation}
where $N_n$ stands for the multivariate normal distribution of dimension $n$, $\vect{I}_n$ is an $n\times n$ identity matrix, $\Xo$ is a full-rank $n\times q$ matrix representing possible nuisance covariates (including a column of ones which corresponds to the constant parameter), $\Xl$ is an $n\times p$ design matrix including the covariates of interest additionally to the ones included in $M_0$, $\al$ is a $q$-dimensional vector of parameters of the null model, $\be$ is a $p$-dimensional vector of regression coefficients related to the covariates of $\Xl$ additionally to the ones included in the null model, and $\tau^{-1}$ is the error variance.

	The null model $(M_0)$ is nested within $M_1$ and represents the model with no effect of the covariates involved in $\Xl$.
Note that this is a generalization of the commonly considered as $M_0$, the constant model (without any covariates) which corresponds to setting $\Xo=\vect{1}_{n\times 1}$ in \eqref{eq:2.1}. 

For the parameterization \eqref{eq:2.1}, the commonly employed conjugate normal-inverse-gamma prior ($NIG$, \citet{ohagan2004}) is formulated as
\begin{align*}
\vect{(\alpha,\beta)}|\tau &\sim N_{q+p}\left(\vect{0},\vect{V} \tau^{-1}\right).
\end{align*}
Assuming prior independence between $\alpha$ and $\beta$, then:
\begin{align*}
	\al|\tau &\sim N_q\left(\vect{0},\vect{V_{\alpha}} \, \tau^{-1} \right) \\
	\be|\tau &\sim N_p\left(\vect{0},\vect{V_{\beta}}  \, \tau^{-1} \right) \\
	\tau &\sim Gamma(k_1/2,k_2/2) 
\end{align*}
where $\vect{V_{\alpha}}$ and $\vect{V_{\beta}}$ are some positive definite matrices, and the gamma distribution is in the shape-rate parameterization. A limiting case of the $NIG$ prior is Zellner's $g$-prior \citep{zellner1986}, as modified and implemented by \citet{liang2008}, which results to 
\begin{align*}
\Xo=\vect{1}&, ~ \al = \alpha   \\
p(\al,\tau) &\propto \tau^{-1} \\
\be|\tau &\sim N_n\left(\vect{0}, g(\Xl^t\Xl)^{-1} \tau^{-1} \right)
\end{align*}
by imposing $\vect{V_{\alpha}}\rightarrow 0$ and $k_1,k_2\rightarrow 0$; where  $\alpha$ is now a scalar. \citet{liang2008} have shown that for the $g$-prior setup:
\begin{align}\label{eq:2.2}
2\log BF_{10}=(n-p-1)\log (1+g) - (n-1)\log\big[ 1+g(1-R_1^2)\big]
\end{align}
where $R_1^2$ is the coefficient of determination for model $M_1$. For the model \eqref{eq:2.1} with $q$ nuisance covariates, the resulting $BF_{10}$ is given by

\begin{align}\label{eq:2.21}
2\log BF_{10}=(n-p-1)\log (1+g) - (n-q-1)\log\big[ 1+g\frac{1-R_1^2}{1-R_0^2}\big]
\end{align}
where $R_0^2$ the coefficient of determination for model $M_0$.

\subsection{Bayes Factors with $g$-priors via test statistics}
Let us assume a model $M_1$ with parameters $\theta=(\theta_0, \theta_1) $ and a sub-model $M_0$ with $\theta_1=0$.
Then, the likelihood ratio statistic for a null hypothesis 
$H_0: \theta_1=0, \theta_0 \in \Theta_0$ vs. the alternative 
$H_1: \theta \in \Theta= \Theta_0 \times \Theta_1$ (which correspond to $M_0$ and $M_1$ as defined above) is defined as 
\begin{align*}
\Lambda_{10}=2\log\left[{\frac {\sup _{\theta \in \Theta_{1}}   f( \y | \theta_0, \theta_1  , M_1 )}
	                           {\sup _{\theta_0 \in \Theta_{0} }f( \y | \theta_0, \theta_1=0, M_0 )}}\right]\,. 
\end{align*}
Commonly, $\Lambda_{10}$ is represented as a function of $\hat{\theta}$, the MLE of $\theta$,  and hence it can be expressed as 
\begin{align*}
\Lambda_{10}=-2\left[\ell_0(\hat{\theta}_0)-\ell_1(\hat{\theta})\right]
\end{align*}
with $\ell_k(\cdot)= \log f( \y | \cdot, M_k )$ denoting the log-likelihood function. This is a functional form which stresses the difference with $BF$s;
in order to calculate the BF one has to integrate the likelihood function over the respective prior distribution for each model, while $\Lambda$ calls for conditional evaluation of the likelihood function given the MLE. Note also that $\Lambda$ is defined for nested hypotheses (or models) and thus $\Theta_0 \subseteq \Theta$ (or equivalently $M_0 \subseteq M_1$). For a comprehensive overview concerning the relationship as well as approximations between $BF$s and $\Lambda$ see \citet{kass1995}. 

Equations \eqref{eq:2.2}  \& \eqref{eq:2.21} make clear that in the case of the g-prior, for the linear model described in \eqref{eq:2.1} (but also general conjugate analysis using the NIG priors as shown in \citet{zhou2018}), the $BF$ comparing the two models is also merely a statistic, i.e. a function of the data only. Considering that $\Lambda_\ell=-n\log(1-R_\ell^2)$ for a model $M_\ell$ \citep{wang2017}, equation \eqref{eq:2.2} shows that for linear regression and $g$-priors, the $BF$ is a simple transformation of $\Lambda$. Furthermore, for the case of simple linear regression (and thus $p$=1), hypothesis in \eqref{eq:2.1} can also be tested via the $T_{\nu}$ statistic with $\nu=n-2$ degrees of freedom for which we have \citep{engle1984}:
\begin{align*}
T^2_{\nu}=\nu\left[\exp \left(\frac{\Lambda}{n}\right)-1 \right] \,.  
\end{align*}
It is straightforward to show that for $p=1$:
\begin{align}\label{eq:2.5}
2\log BF_{10} = (n-2)\log (1+g) - (n-1)\log \left[ 1+g \left(\frac{T^2_{n-2}}{n-2}+1\right)^{-1}\right]\,. 
\end{align}
Thus for simple linear regression and the $g$-priors setting, the resulting $BF$ can be expressed analytically as a function  of the $T_{\nu}$ statistic. 
Therefore, by combining individual $T$ statistics,  we can eventually also obtain the overall Bayesian evidence in the form of a $BF$.

\section{Bayesian Evidence Synthesis}
\subsection{General Setting and notation}
 Suppose we have $K$ studies each resulting from the same data generating model, i.e. the common effect model as defined above. The sample size of each study is denoted by $n_k$ and the size of the combined dataset is given by  $N=\sum_{k=1}^K n_k$. 
 Here we focus on the "simple" regression case where $p=1$ and ${X_1}$ can be either a continuous scalar covariate or 
a binary factor. The latter represents the case of a simple $t$-test comparing two groups, probably the  most common problem encountered in the meta-analytic context. 
We assume that each study $k$ with response data $\y_k$ provides a test statistic $\mathcal{T}_k$ suitable to evaluate the (same) null hypothesis $H_0$. This can be a Student $t$-test statistic $T_k$ (i.e. the $t$ statistic
 from study $k$ with $\nu_k=n_k-2$ degrees of freedom), a likelihood ratio statistic  $\Lambda_k$, or, a Bayes factor  $BF_{10{k}}=BF_{k}$. 
 We will refer to a $T_k$ statistic as a \textit{directional} statistic, as it retains information about
the sign of the effect, while $\Lambda$ and $BF$ are \textit{undirectional} (with respect to the null and alternative defined in \eqref{eq:2.1}). 
 
 Furthermore, we will denote by $\widehat{\beta}_k$ the absolute  estimated effect from study $k$. 
Note that for each of the ${T}_k=\widehat{\beta}_k/{SE}(\widehat{\beta}_k)$, 
it holds that ${T}_k \sim t_{n_k-2}(\sqrt{ss_k} \beta^{sd}_k)$, where $t_{\nu}(c)$ a $t$ distribution with $\nu$ degrees of freedom and non-centrality parameter $c$, 
$ss_k=(n_k-1)s_k^2$ with $s_k^2$ denoting the sample variance of $X_1$ in dataset $k$
and $\beta^{sd}=\beta\tau^{-\frac{1}{2}}$ is the corresponding standardized, common across studies, effect size parameter; reminder: $\tau^{-1}$ is the error variance in the regression model.
For the case where $X_1$ is a binary factor (i.e. we have a two-sample $t$-test), 
then $ss_k={\frac{n_{k(1)}n_{k(2)}}{n_{k(1)}+n_{k(2)}}}$, 
where $n_{k(j)}$ is the sample size of group $j$ at study $k$.

Each $\mathcal{T}_k$ is comparing the same models $M_0$ and $M_1$, with the latter describing the addition of one covariate to $M_0$. 
In what follows, we will stick to the case where $M_0$ is the null model which assumes a constant expected response across all observations
 and $M_1$ describes either a difference in means between two groups for a dichotomous ${X_1}$, or the linear effect of a continuous ${X_1}$
 on the response $Y$. 
The proposed methodology can be easily employed to accommodate the synthesis of effect estimates in cases they are extracted from model comparisons within the multiple linear regression setup, as discussed in Section \ref{sec:2.2} (see e.g. Equation \ref{eq:2.21}).

We  denote by $BF$, $\Lambda$, $T$ the true, unknown, values of the statistics that we would obtain if all raw data were available, $BF_k$, $\Lambda_k$, $T_k$ are the corresponding statistics calculated from the $k$-th dataset and $\widetilde{BF}$, $\widetilde{\Lambda}$, $\widetilde{T}$ denote the synthesized statistics from the respective sample specific test statistics $\{BF_k\}_{k=1}^K$, $\{\Lambda_k\}_{k=1}^K$ or  $\{T_k\}_{k=1}^K$. 
Our interest lies in calculating $BF_{10}=BF$, i.e. the evidence taking into account all $K$ datasets. 
Thus, we are interested in methods that result in $\widetilde{BF}\approx BF$.  

\subsection{The product of Bayesian evidence}
The product of BFs $\widetilde{BF} =\prod_{k=1}^K BF_k$ can deliver evidence for the fixed effects model \citep{van_lissa_tutorial_2024}, \cite{van_wonderen_bayesian_2024}), but is not appropriate for the, usually assumed, common effect model.  
This approach suggests that the marginal likelihood of the full dataset can be obtained as the product of the marginal likelihoods of each sub-sample and thus $\widehat{m}^{(K)}=\prod_{k=1}^{K} m_{k}$ where $\widehat{m}^{(K)}$ denotes the estimate of the marginal likelihood of the whole sample based on the $m_{k}$ marginal likelihoods of the $K$ sub-samples. This is not true since (ignoring parameters and denoting data by $\vect{y}$)
\begin{align*}
m^{(K)}&=f(\vect{y}^{(K)})=f(\vect{y}_{1},\vect{y}_{1},\dots,\vect{y}_{K}) \\
&=f(\vect{y}_{1})\times f(\vect{y}_{2}|\vect{y}_{1})\times \dots \times f(\vect{y}_{K-1}|\vect{y}_{K-2},\dots,\vect{y}_{1}) \times f(\vect{y}_{K}|\vect{y}_{K-1},\dots,\vect{y}_{1}) \\
&= \prod_{k=1}^{K} f(\vect{y}_{k}|\vect{h}_{k})
\end{align*}
where $\vect{h}_{k}$ is the history of sub-sample $k$, i.e. the sub-samples incorporated to the analysis previously to $\vect{y}_{k}$, given by $\vect{h}_{k}=\left(\vect{y}_{1},\dots,\vect{y}_{k-1}\right)$ and $\vect{h}_{1}=\emptyset$. 
Here, we assume exchangeability between studies, which means that the ordering of the samples does not matter for these calculations.
Thus, the time ordering will not influence our model and the association between $Y$ and $X_1$.  
Therefore the combined marginal likelihood $m^{(K)}$ is the product of the marginal likelihoods, which are the \textit{conditional} 
marginal likelihoods of $\vect{y}_{k}$ given the previous history $\vect{h}_{k}$, given by
\begin{align*}
f(\vect{y}_{k}|\vect{h}_{k})=\int f(\vect{y}_{k}|\vect{\theta}) p(\vect{\theta}|\vect{h}_{k}) d \theta
\end{align*}
where $p(\vect{\theta}|\vect{h}^{(k)})$ the posterior distribution of the relevant parameters for the 
data $\vect{y}^{(1)},\dots,\vect{y}^{(k-1)}$. The computation of this expression requires the full set of raw data. 
A demonstrative example of why the product of $BF_k$s resulting from partitioning a dataset is not the same as the $BF$ computed from the pooled data is provided in \citet{rouder2011}.

\subsection{Bayesian Evidence Synthesis using Cauchy priors}
\label{sec:JZS}

 Bayesian $t$-tests employing $BF$s have received attention mainly in the quantitative psychology literature; 
 see for example  \citet{gonen2005}, \citet{rouder2009} and \citet{gronau2019}. See \citet{heck2023review} for a recent review. 
 
An alternative to the conjugate prior setup and the g-prior specification for testing \eqref{eq:2.1}, 
is to use the a standard Cauchy prior on $\beta^{sd}$ as originally suggested by \citet{zellner1980}.  
The Zellner-Siow prior is frequently combined with a Jeffreys prior for $\tau$ which sometimes is referred to as the Jeffreys-Zellner-Siow prior (JZS) \citep{rouder2009}.
This prior can be considered as a mixture of $g$-priors  \citep{liang2008} since it can be written in the following  hierarchical manner: 
 $$
 \beta|\sigma^2_{\beta} \sim N(0,\sigma^2_{\beta}) \mbox{~~and~~} \sigma^2_{\beta}\sim \mbox{Inv-}\chi^2_1.
 $$  

\citet{rouder2009} adopt a slightly different parameterization for the two sample $t$-test (originally suggested in \citep{gonen2005}) so that inference is based directly on the $\beta^{sd}$ parameter as described here. This parameterization essentially translates to the $X_1$ variable being centered in the univariable regression setup \citep{wetzels2012}. 
We may write the JZS Bayes factor as a function of the $T=T_{\nu}$ statistic, by the following expression \citep{gronau2019}
\begin{align*}
BF_{10} = \frac{\int_{-\infty}^{+\infty} f_{\nu}(T; \sqrt{ss_{x}} \beta^{sd}) g(\beta^{sd})d \beta^{sd} }{f_{\nu}(T;0)} 
\end{align*}
where $f_{\nu}(\cdot;c)$ the pdf of a $t$-distributed random variable with $\nu$ degrees of freedom and non-centrality 
parameter $c$ and $g(\cdot)=f_{1}(\cdot;0)$ the pdf of a standard Cauchy distribution. The number of degrees of freedom depends on whether one- or two-
sample $t$-tests are employed ($n-1$ or $n-2$ respectively).
 
\citet{rouder2011} suggest treating the individual $T_k$ statistics that arise from different studies as a sample of data points from respective $t$ distributions. With this formulation the only parameter in both models is $\beta^{sd}$ which is 0 under $H_0$ and follows a standard Cauchy distribution under $H_1$. 
Based on that formulation, a meta-analytic $BF$ is suggested (here generalized for the univariable regression setting):

\begin{align*}
\widetilde{BF}_{10} = \frac{\int_{-\infty}^{+\infty} \prod_{k=1}^{K} f_{\nu_k}(T_k; \sqrt{ss_{k}} \beta^{sd}) g(\beta^{sd}) d \beta^{sd}}{\prod_{k=1}^{K} f_{\nu_k}(T_k;0)}. 
\end{align*}
This method for the synthesis of Bayesian evidence will be abbreviated in the following sections as 
\textit{Meta-$BF_{JZS}$}.

\section{Proposed methods: Bayes Factors Synthesis under the $g$-priors}

\subsection{Evidence synthesis and availability of data}
In order to calculate \textit{Meta-BF$_{JZS}$}, one needs to have access to $T_k$, $n_k$ and $ss_{k}$ for all studies. 
However, following a systematic review, the level of detail of the available information concerning the effect of a variable can vary across different studies \citep{higgins2019}. 
For example, it is not uncommon that information concerning $ss_{k}$, i.e. the sum of squares of the covariate $X_1$, is not available in all studies. 
Another possibility is that, in some studies, only the effect estimates are presented in the form of 
$\widehat{\beta}_k$ but not the observed values of the test statistics $T_k$ (thus with unknown ${SE}(\widehat{\beta}_k)$).
 
We therefore present methods for Bayesian evidence synthesis for three scenaria of minimum data availability. We employ the terms directional/undirectional and one-sided/two-sided interchangeably, since they coincide for the case of testing the effect of a single covariate. 

\begin{itemize}
\item \textbf{Case 1 - Detailed Information (D)}: In this scenario, one- or two- sided  $\mathcal{T}_k$ 
($T$, $\Lambda$, $T^2$, $p$-value, or $BF_k$), $n_k$, effect direction and $ss_{k}$  are available from all $K$ studies. 
This scenario is reasonable to expect when considering evidence synthesis for the $t$-test case as it simply requires the sample size per group to be reported. For the simple linear regression case, it requires to record the observed sum of squares $ss_k$ of the covariate $X_1$ for each study $k$.

\item \textbf{Case 2 - Partial Information (P)}: One- or two- sided  $\mathcal{T}_k$ ($T$, $\Lambda$, $T^2$, $p$-value, or $BF_k$), 
$n_k$ and effect direction available. For binary factors $X_1$ this scenario translates to unknown sample sizes per group. 
For continuous covariates $X_1$, this corresponds to unreported $ss_k$ - which is a common scenario \citep{becker2007}.

\item \textbf{Case 3 - Limited Information (L)}: Only two-sided $\mathcal{T}_k$ ($\Lambda$, $T^2$, two-sided $p$-value, or $BF_k$) 
and $n_k$ available without knowing the direction of the effect.
 In this, extreme, scenario, there is no information on the direction of the effect from all the $K$ datasets and effect measures are limited to an 
undirectional statistic \citep{cucherat2000}. 

\end{itemize}

Below we present methods for calculating $g$-priors based $\widetilde{BF}$ for the above varying levels of available information. 
We present an alternative for $JZS$-based $\widetilde{BF}$ when the same level of detail is obtainable (Case 1- D), but also for the 
scenaria with more restricted access to information regarding the effect under study (Cases 2 \& 3). 

As with \textit{Meta-BF$_{JZS}$}, we approach the problem through the prism of synthesizing the frequentist test 
statistics (mainly $T_k$) and then transform the combined evidence into $\widetilde{BF}$. This serves a pragmatic purpose, 
since most published studies employ the classical statistical approach, but it also demonstrates how the methods can be used 
for translating frequentist evidence into Bayesian. We, therefore, aim at arriving at a $\widetilde{BF}_g$ (as in: $g$-prior based
 meta analytic $BF$), by synthesizing the $K$ available evidence measures. As all the statistics discussed here 
($\Lambda_k$, $T_k^2$, two-sided $p$-value, $BF_k$) can be transformed to a $T^2$ statistic, we discuss below methods for synthesizing
 those $T^2_k$ statistics into a $\widetilde{BF}_g$.
 
Note that, the general problem presented in \eqref{eq:2.1} is a two-sided hypothesis test.  Evidence synthesis on the basis of 
undirectional statistics (like the ones mentioned in Case 3) comes with additional complications. Specifically, two-sided statistics 
are not one-to-one transformations of the underlying one-sided statistics, and thus evidence synthesis methods might fall short \citep{held2018}. As an illustration, consider a standard $t$-test and a set of data that arises from $H_0$, that is, the true 
$\beta=0$. A random split of the data in $K$=2 partitions could result in $\widehat{\beta_1}>0$ and $\widehat{\beta_2}<0$ , 
or vice-versa. Depending on the magnitude of the sub-samples’ deviations from $H_0$, this in turn could translate to a pair 
of small two-sided $p$-values or, equivalently, large test statistics (i.e. $T^2$, $\Lambda$ or $BF$). Synthesizing any of these 
two-sided statistics without taking into account the sign of $\beta_k$ would result to a misleading combined estimate. 
Therefore, for Cases 1 \& 2 evidence synthesis will be on the basis of the $T_k$ statistics, while for Case 3, we will employ 
two-sided $p$-value combination methods. 

\subsection{Methods for Detailed information (D) }
\label{sec:Full_info} 

In this section we deal with the full information scenario where the study specific original test statistics $T_k$ are 
directly available or can be obtained from other reported test statistics $\mathcal{T}_k$ and the signs of $\hat{\beta}_k$. 

By employing a technique that involves a normal approximation for the $T_k$ statistics, the overall $\widetilde{T}$ can be estimated by 
\begin{equation}\label{eq:2.6}
\widetilde{T}=\sum_{k=1}^K w_k \left[ \left(\mathds{1}_{\{ \hat{\beta}_k>0 \}} -\mathds{1}_{\{\hat{\beta}_k<0 \}} \right) \times \sqrt{T^2_k} \right] 
\end{equation}
where $\mathds{1}_{\{ \cdot \}}$ the indicator function. Note that we present results on the basis of $\sqrt{T^2_k}=|T_k|$ in order to make the connection with $BF_k$ obvious.
This meta-analytic test statistic is based on the inverse normal method, commonly employed in the field of adaptive trial design where test statistics from different stages are combined \citep{lehmacher1999}. Note that for known and equal variance across studies, and weights that equal $w_k=\sqrt{n_k/N}$, then - by abusing notation and referring by $T$ to the corresponding $Z$ statistic 
- $\widetilde{T}=T$. 
 
By combining \eqref{eq:2.5} \& \eqref{eq:2.6} one can obtain $\widetilde{BF}$ by utilizing either $T^2_k$ or $BF_k$. Weights $w_k$ should be chosen appropriately and satisfy 
$\sum_{k=1}^K w_k^2=1$ \citep{lehmacher1999}. By incorporating inverse variance weighting and taking into account the relative variances of the $T_k$ statistics,
 the weights are given by  $w_k=\sqrt{\frac{v_k^{-1}}{\sum_{k=1}^K v_k^{-1}}}$  where:
\begin{align}\label{eq:2.7}
v_k &= H(\nu_k/2)^{-2}\frac{\nu_k}{ss_{k}(\nu_k-2)}+\left[\left(\frac{\nu_k}{\nu_k-2}\right)H(\nu_k/2)^{-2}-1 \right]d_k^2 \\[1em] 
\mbox{with~~} 
d_k &= \frac{{T}_k}{H(\nu_k/2)\sqrt{ss_{k}}} \mbox{~~and~~} 
H(z)=\sqrt{z}\left[\frac{\Gamma(z-1/2)}{\Gamma(z)}\right]. \nonumber 
\end{align}
These weights are the result of combining Equation 23 from \citet{malzahn2000}, and the requirement for $\sum_{k=1}^K w_k^2=1$. 
They essentially apply inverse-variance weighting for the individual $T_k$ statistics and treats their sum as a sum of
independent $N(\sqrt{ss_k} \beta^{sd}_k,1)$ variables. 
In the following, we will refer to this meta-analytic estimator as \textit{Meta-$BF_{g}^{(D)}$}.

\subsection{Methods for Partial Information (P) }\label{sec:Partial_info} 

In this section we deal with the more realistic scenario where partial information is available. 
Hence, if the variance component $ss_k$ is not available for every study then we cannot implement neither the 
Meta-$BF_{g}^{(D)}$ approach proposed in Section \ref{sec:Full_info} nor the $JZS$ based $\widetilde{BF}$  of \citet{rouder2009}  presented in Section \ref{sec:JZS}. 
In our proposed Meta-$BF_{g}^{(D)}$, the problem is mainly that we cannot obtain the weight $w_k$ 
of each study $k$ since \eqref{eq:2.6} requires the values of the corresponding sum of squares $ss_k=(n_k-1)s_k^2$. 

 In such cases, a simple but reasonable solution is setting $w_k=\omega_k=\sqrt{n_k/N}$ in equation \ref{eq:2.7} and then 
employ a normal approximation-based inverse variance weighing for all the $T_k$ statistics. 
 This is equivalent to adopting the assumption that all variances are known and equal across the $K$ studies. 
This approach will be abbreviated as \textit{Meta-$BF_{g}^{(P)}$}.

\subsection{Methods for Limited Information (L) }\label{sec:Limited_info}

Finally, we consider the case where the available information is restricted to $p$-values or test statistics where the direction of the effect is not evident. Hence, if  we only have access to two-sided $p$-values then we can think of using any $p$-value combination method for the synthesis of frequentist evidence, and then transform the result to a Meta-BF ($\widetilde{BF}$). 
In order to synthesize a vector of  $p$-values $\vect{p}=\{p_1,\dots,p_K\}$ testing the same $H_0$ vs. $H_1$ coming from $K$ independent studies the result that  
\begin{equation} 
p_i|H_0\sim Uniform(0,1) ~~\forall~i\in \{1\dots K\} \label{p_values_uni}
\end{equation} 
is of fundamental value. 
By using \eqref{p_values_uni}, several combination methods exist in order to obtain an overall p-value  $\widetilde{p}=f(\vect{p})$ 
which will act as a global measure of departure from $H_0$. 
Two of the most common statistics for combining $p$-values, for which the distribution under 
the null hypothesis is available are given by:

(a) the Fisher's method \citep{fisher1934}, and 
(b) the Stouffer's method \citep{stouffer1949}. 
The $p$-value based statistics for these two methods are given by 
\begin{align}
\textrm{Fisher's method: } S_F&=-2\sum_{k=1}^K \log(p_k) \sim \chi^2_{2K} \, ,  \label{Fishers} \\
\textrm{Stouffer's method: } S_S&=\sum_{k=1}^K \Phi^{-1}(1-p_k) \sim N(0,K). \label{Stouffers} 
\end{align}
The following strategy could be adopted in order to obtain the Meta-BF,
 $\widetilde{BF}$, from a (two-sided only) $\mathcal{T}_k$: 
\begin{enumerate}
\item For each $\mathcal{T}_k$, calculate the associated (two-sided) $p$-value; 
\item Synthesize the resulting vector $\vect{p}=\{p_1,\dots,p_K\}$ via \eqref{Fishers} or \eqref{Stouffers} and obtain $\widetilde{p}$; 
\item Transform $\widetilde{p}$ to an appropriate $\widetilde{\mathcal{T}}_k$.
\item Obtain $\widetilde{BF}$ from $\widetilde{\mathcal{T}}_k$ using \eqref{eq:2.5}. 
\end{enumerate}
Such an approach would suffer from the loss of directionality in each $p_{k}$.

\citet{owen2009} addresses this issue by exploring a long-ignored method for $p$-value combination dating 
back to \citet{pearson1934}. It can be briefly described as running a Fisher or Stouffer style test 
for left sided alternatives and one for right-sided alternatives and then considering the most extreme of the two. 
In other words, for a set of study specific two-sided only $p$-values $\vect{p}=\{p_1,\dots,p_K\}$, and by
 implementing the Stouffer's method, we then consider:
\begin{equation} \label{eq:2.55} 
\begin{aligned} 
S_S^L&=\sum_{k=1}^K \Phi^{-1}(1-p_k/2),~~ 
S_S^R&=\sum_{k=1}^K \Phi^{-1}(p_k/2), \mbox{~~and~~} 
S_S^C&=\max( S_S^R,S_S^L).
\end{aligned}
\end{equation}
Note that $S_S^R=-S_S^L$ and $S_S^C=|S_S^L|=|S_S^R|$.
This $p$-value based statistic can be adopted as a combination method if no information on the direction of the $\beta_k$ is 
available from the $K$ studies. A modified weighted version of \eqref{eq:2.55} could be employed with 
$S_S^C=\sum_{k=1}^K \omega_k |T_k|$ as an approximation to $\widetilde{T}$ which can be transformed to 
a $\widetilde{BF}$ via \eqref{eq:2.5}.

We will refer to this method as \textit{Meta-$BF_{g}^{(L)}$}. The methods described alongside the required ingredients for their implementation and abbreviations used in what follows 
are summarized in Table \ref{tab:sum}.

\begin{table}[b!] 	
	\centering
\def\arraystretch{1.5}	
		\begin{tabular}{lccc}
& \multicolumn{3}{c}{Minimum data requirements per study} \\
\hline
Two sided only evidence ($\Lambda$, $T^2$, $p$-value, or $BF_k$)&$\checkmark$&$\checkmark$&$\checkmark$\\
Sample size per study ($n_k$) &$\checkmark$&$\checkmark$&$\checkmark$ \\
Effect direction ($sign(\hat{\beta}_k$))&$\checkmark$&$\checkmark$&\\
Sum of squares of covariate (${ss_{k}}$)&$\checkmark$&&\\
\hline
\multirow{2}{*}{Method (Section)} & \textit{Meta-$BF_{JZS}$} (\ref{sec:JZS}) & {\textit{Meta-$BF_{g}^{(P)}$} }& {\textit{Meta-$BF_{g}^{(L)}$} }\\
& \textit{Meta-$BF_{g}^{(D)}$} (\ref{sec:Full_info})&(\ref{sec:Partial_info})&(\ref{sec:Limited_info})\\ 
	\end{tabular}
	\caption{Summary of methods for Bayesian evidence synthesis with respect to the available data from included studies in a meta analysis}
	\label{tab:sum}
	\end{table}

\section{Simulation Study}
In this section we compare the performance of the discussed methods through two simulation studies. In the first the total sample size $N$ is retained fixed. The main goal is
 to assess whether the meta-analytic $\widetilde{BF}s$ are
providing the same evidence as the $BF$ obtained from the
full data.

The second simulation follows the standard simulation approach
typically employed for the comparison of meta-analytic methods, 
where the total and within-study sample size is random \citep{pateras2018}. The main goal of this exercise 
is to compare the operational characteristics of the approaches
 under study for increasing $K$. 

\subsection{Fixed sample size}
\subsubsection{Simulation Plan}
The data generating model was $Y\sim N(\beta X_1,1)$ and $X_1$ was either a group indicator, representing an independent 
samples $t$-test, or a fixed realization of a $N\left(0,\frac{1}{4}\right)$
 variate, materializing a simple linear regression scenario 
with variance approximately equal to the variance used in
 the $t$-test simulation scenario.
The total sample size was set to $N=1000$ and we explored 
$K=\{2,5,10\}$ partitions of the full data set. 
Group sizes were balanced in the case of the $t$-test. We employ
 two over-arching scenarios for the partitions, one where 
$n_k=N/K\Rightarrow \omega^2_k=1/K$ and thus all sub-samples 
were balanced and one with unequal partition where:
\begin{itemize}
\item For $K=2$, $\omega^2_1=0.7$ and $\omega^2_2=0.3$
\item For $K=5$, $\{\omega^2_k\}_{k=1}^5=\{0.05,0.1,0.15,0.3,0.4)\}$ 
\item For $K=10$, $\omega^2_1=\dots=\omega^2_6=0.05$, $\omega^2_7=0.1$ 
and $\omega^2_8=\omega^2_9=\omega^2_{10}=0.2$. 
\end{itemize}
We will refer to the former as the EQ scenario, and the latter
 as the UNEQ. Concerning the effect size, we assess the null
hypothesis/model and levels of moderate 
departure from the null, i.e. $\beta \in \{0,0.1,0.2,0.3\}$. We 
conducted $R=1000$ replications per scenario.

For each of the $K$ sub-samples per scenario, $BF_k$ and $T_k$ were calculated as well as the $BF$ from the full dataset.
 For the $g$-prior setup we consider $g_k$=$n_k$ and $g=N$.
 As a basic assessment of the performance of each method, we focus on
the quantification of evidence in a discrete sense. According to
 \citet{kass1995}, the evidence provided by a $BF_{10}$ can be 
categorized in four levels of evidence presented in
in Table \ref{tab:1}. Per simulation scenario and for each method,
 a contingency table is constructed 
presenting the cross-classification of the levels of evidence for the
(full dataset) $BF$ and each of the meta-analytic $\widetilde{BF}$ proposed here.
 Agreement in terms of correct classification in the categories
 of strength of evidence is assessed via the weighted version of Cohen’s 
$\kappa$ statistic, using quadratic weights to assign larger importance to larger 
discordances \citep{fleiss1969}. The categories used to evaluate agreement are shown in Table \ref{tab:2}, alongside the colour scale used 
for visualization of the level of agreement. 
\begin{table}[h!] 
	\centering
\def\arraystretch{1.5}	
		\begin{tabular}{ccc}
$2\log BF_{10}$&$BF_{10}$& Evidence against $H_0$\\
\hline
0 - 2	 & 1 - 3 & Not worth more than a bare mention \\
2 - 6	 & 3 - 20 & Positive \\
6 - 10	 & 20 - 150 & Strong\\
$>$10 & $>$150 & Very strong \\
\hline
	\end{tabular}
	\caption{Level of evidence provided by $BF$}
	\label{tab:1}	
	\end{table}

\begin{table}[h!] 
	\centering
\def\arraystretch{1.5}	
		\begin{tabular}{cc}
Cohen's $\kappa$ & Level of Agreement \\
\hline
$<0$ &\textcolor{LightGray}{No agreement}\\
$0-0.2$ &\textcolor{LightGray}{Slight agreement}\\
$0.2-0.4$ &\textcolor{Gray}{ Fair agreement}\\
$0.4-0.6$ & \textcolor{Gray}{Moderate agreement}\\
$0.6-0.8$ & \textcolor{DarkSlateGray}{Substantial agreement}\\
$0.8-1.0$ & \textcolor{Black}{Almost Perfect agreement}\\
\hline
	\end{tabular}
	\caption{Level of agreement suggested by (weighted) Cohen's $\kappa$}
	\label{tab:2}	
	\end{table}		

\subsubsection{Results}\label{sec:res1}
\begin{figure}
\captionsetup[subfigure]{labelformat=empty}

\subfloat[K=2]{\includegraphics[width = .33\textwidth,height=0.24\textheight]{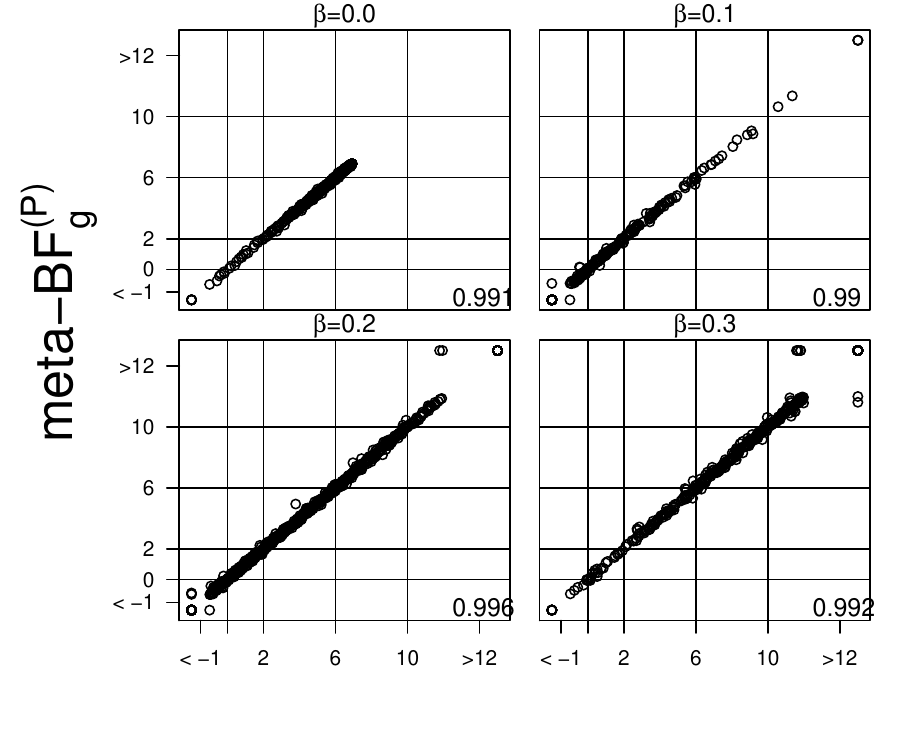} }
\subfloat[K=5]{\includegraphics[width = .33\textwidth,height=0.24\textheight]{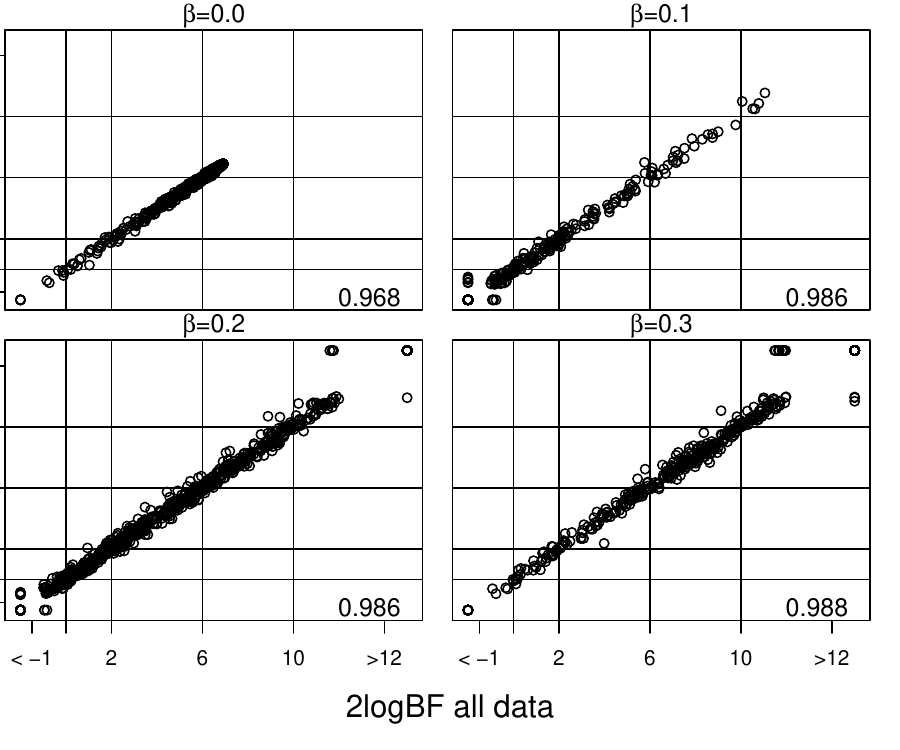} }
\subfloat[K=10]{\includegraphics[width = .33\textwidth,height=0.24\textheight]{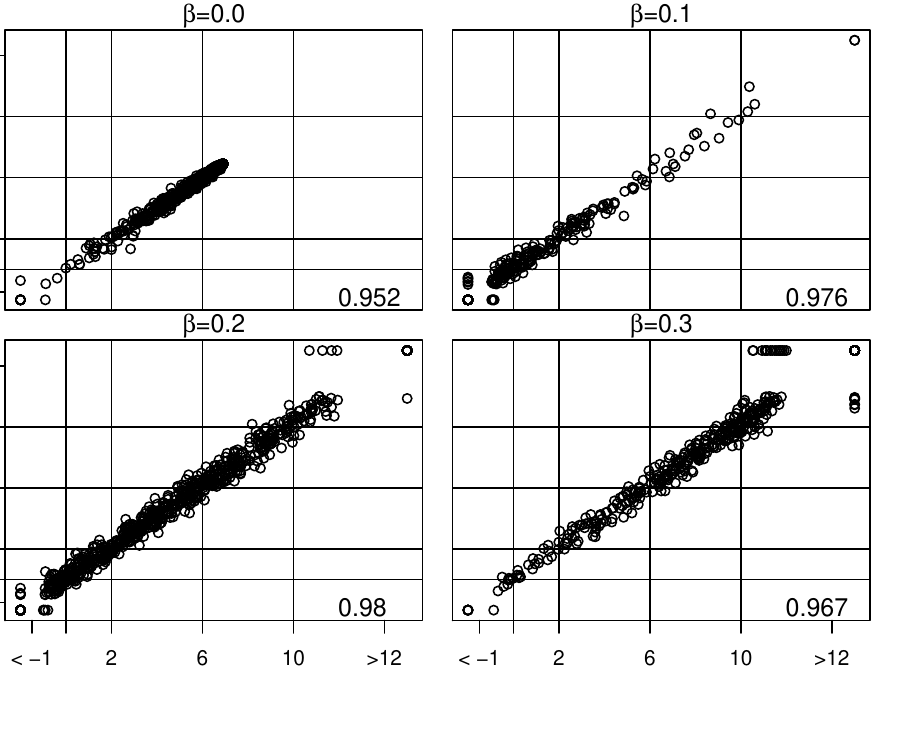} }\vspace{-.5cm}  \\ 
\subfloat{\includegraphics[width = .33\textwidth,height=0.24\textheight]{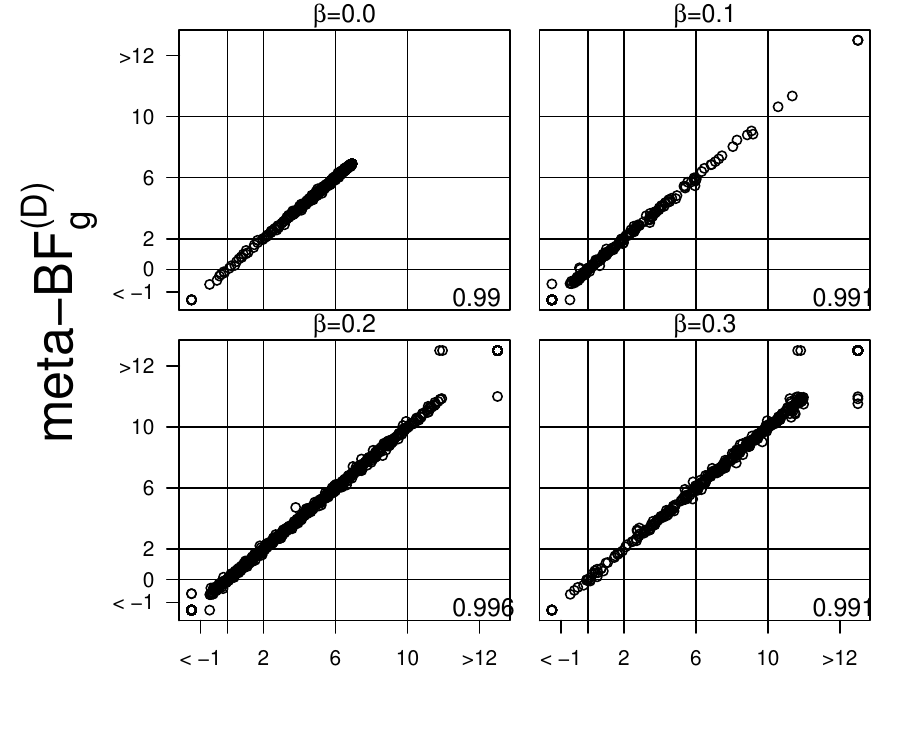} }
\subfloat{\includegraphics[width = .33\textwidth,height=0.24\textheight]{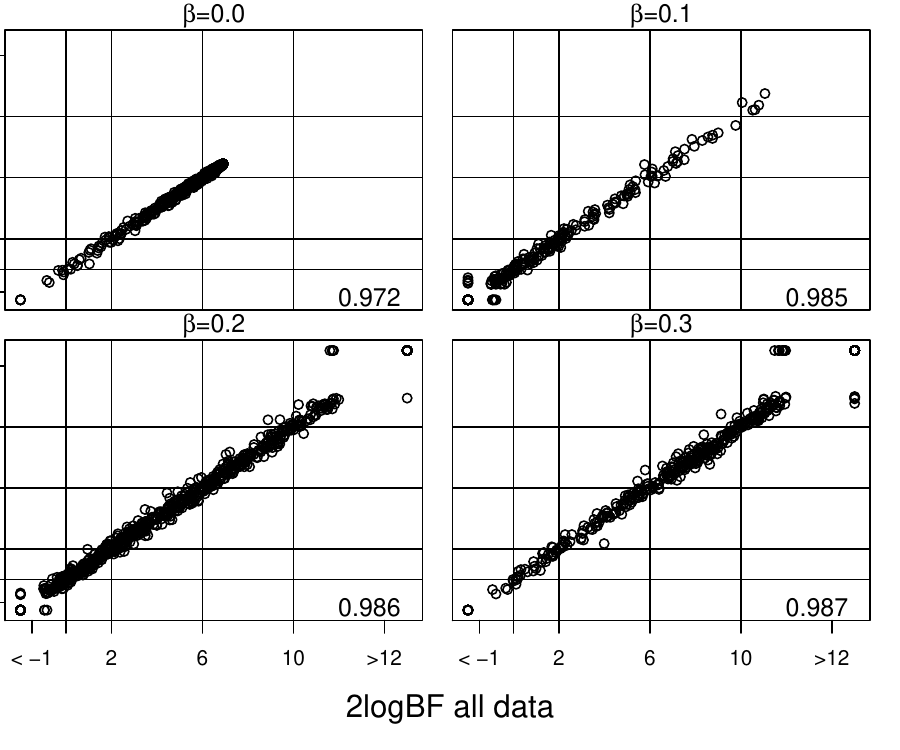}} 
\subfloat{\includegraphics[width = .33\textwidth,height=0.24\textheight]{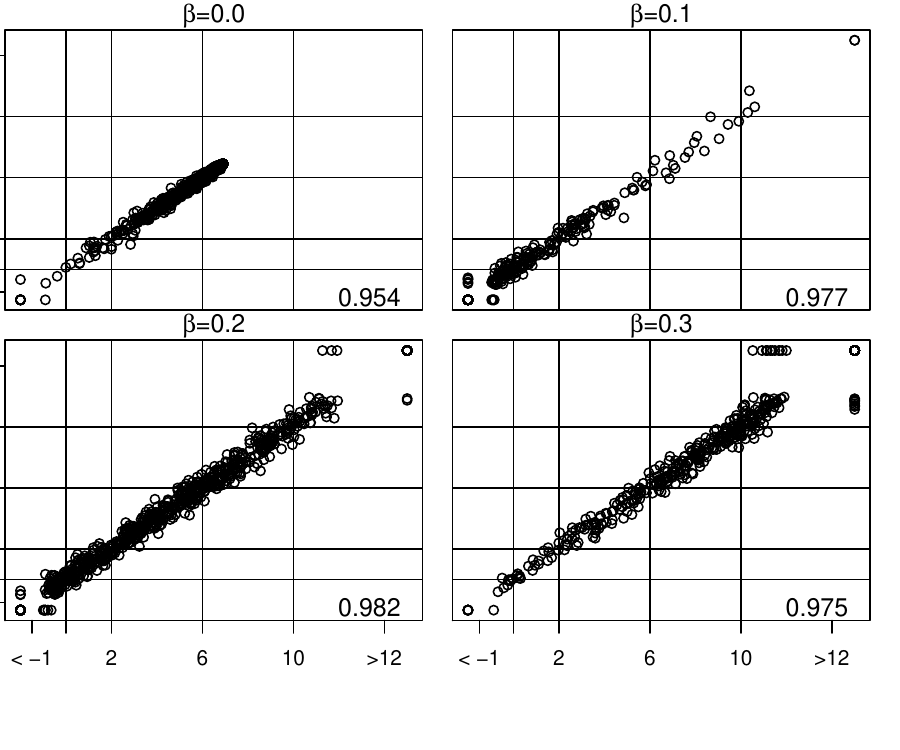}}\vspace{-.5cm} \\ 
\subfloat{\includegraphics[width = .33\textwidth,height=0.24\textheight]{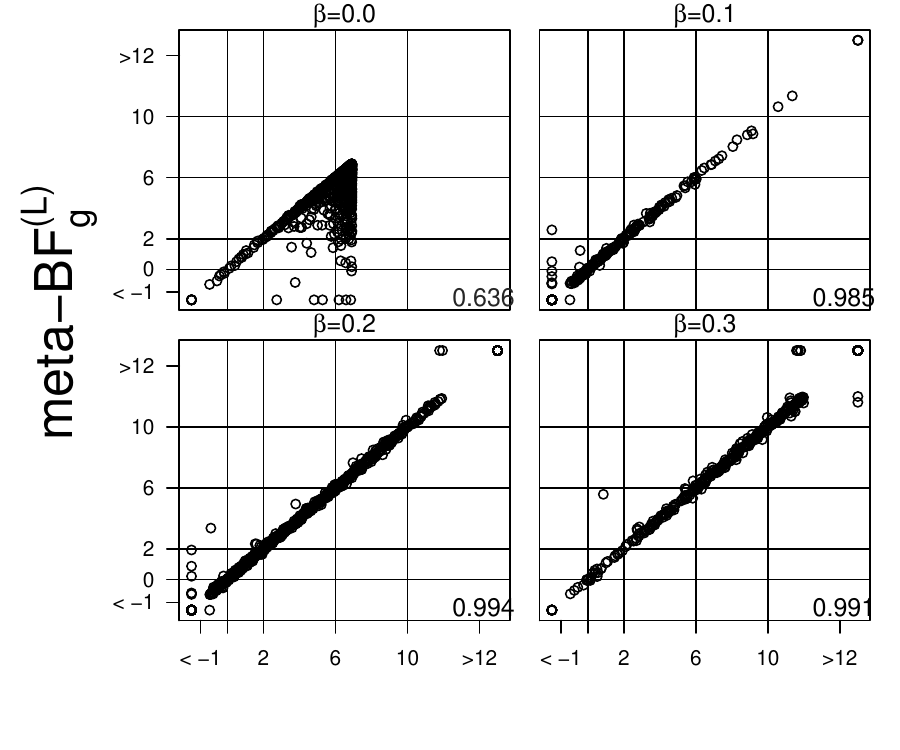} }
\subfloat{\includegraphics[width = .33\textwidth,height=0.24\textheight]{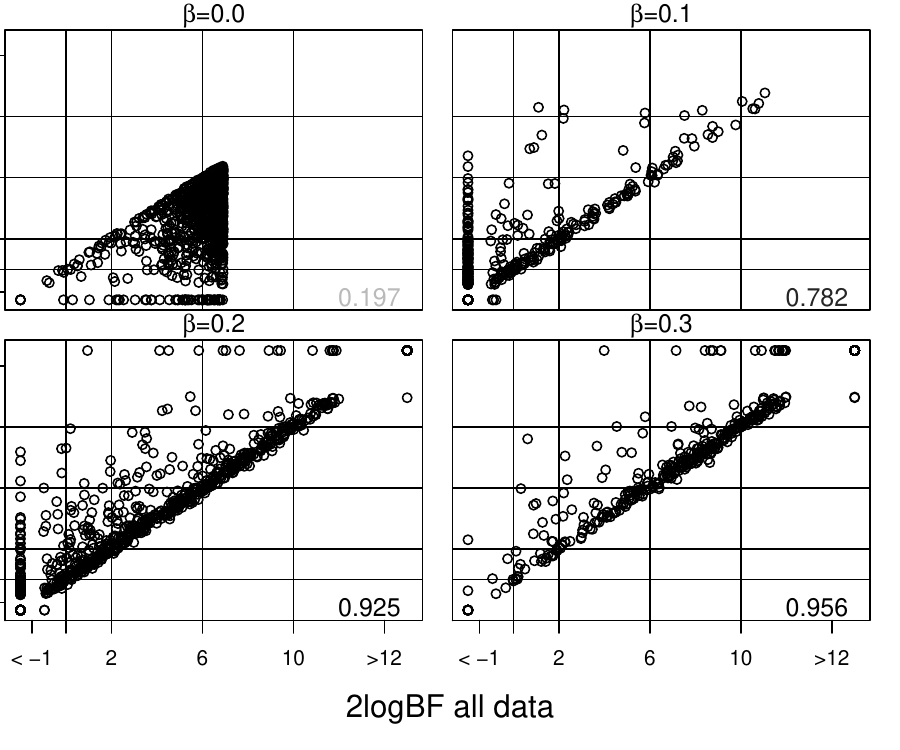}}
\subfloat{\includegraphics[width = .33\textwidth,height=0.24\textheight]{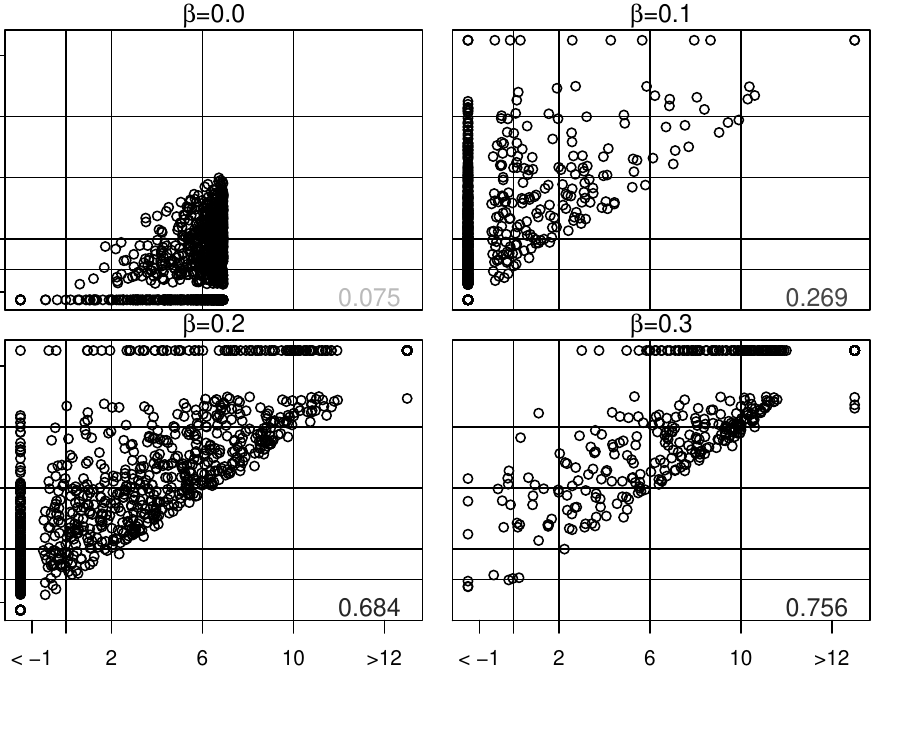} }\vspace{-.5cm} \\ 
\subfloat{\includegraphics[width = .33\textwidth,height=0.24\textheight]{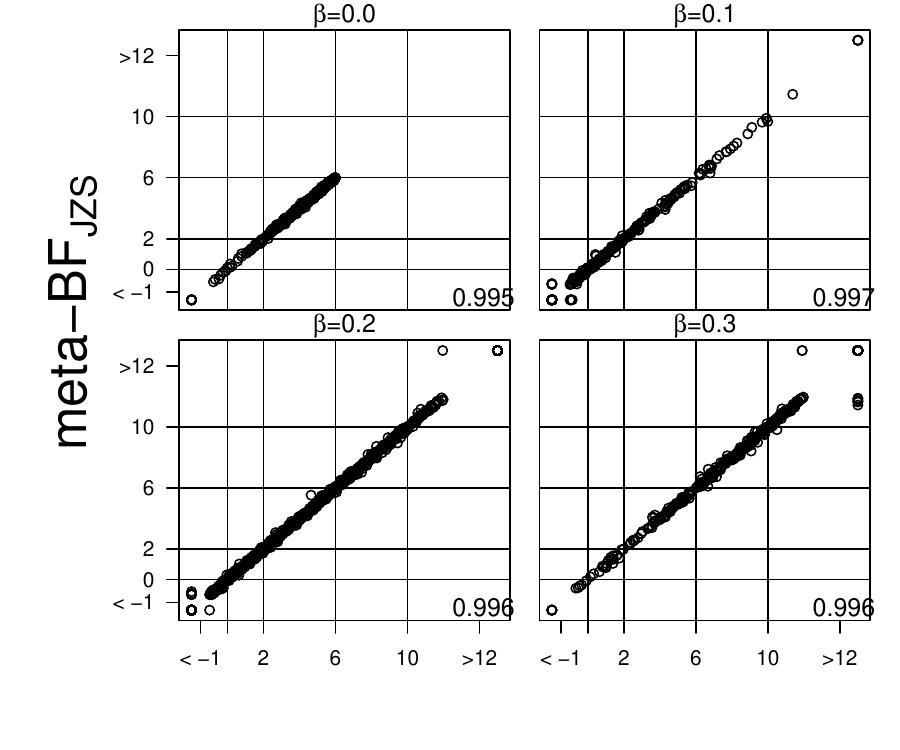} }
\subfloat{\includegraphics[width = .33\textwidth,height=0.24\textheight]{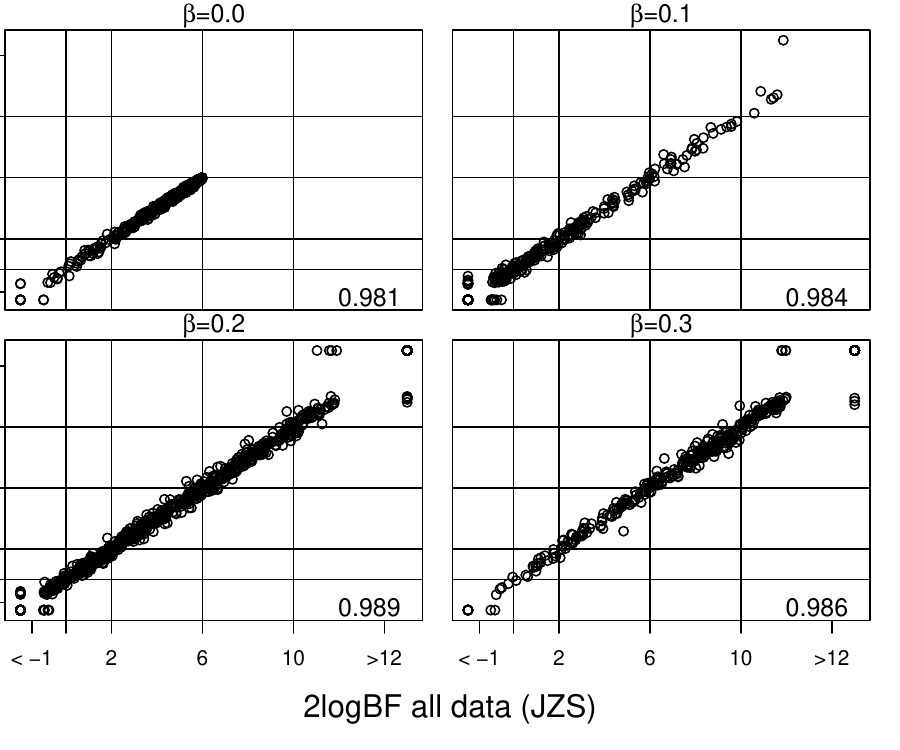}} 
\subfloat{\includegraphics[width = .33\textwidth,height=0.24\textheight]{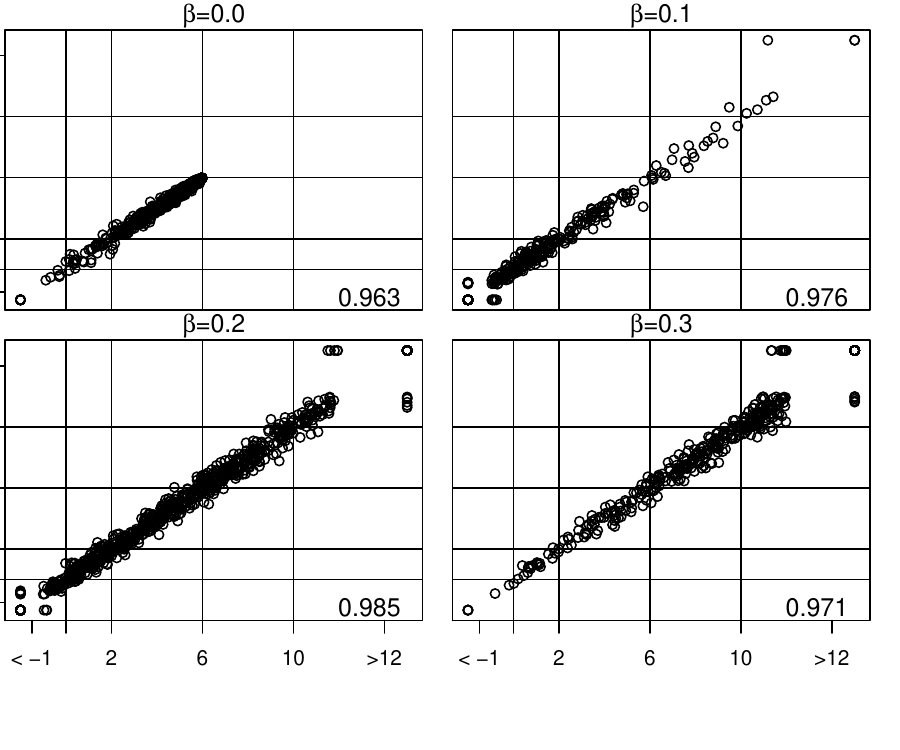} }
\caption{Scatter plots and weighted Cohen's $\kappa$ statistics for the
 agreement between $BF$ and $\widetilde{BF}$ for binary $X$ 
($t$-test) and EQ scenario. $2\log BF_{10}$ are shown, except 
for $\beta=0$ where $2\log BF_{01}$ is shown. Grid lines 
indicate the categories of level of 
evidence as described in Table \ref{tab:1}. Values $<$-1 and $>$ 12 are shown as equal. }
\label{fig:1}
\end{figure}

The results of the first simulation exercise are shown in Figure \ref{fig:1}. It presents the association scatterplots between $2\log BF$
 and $\widetilde{2\log BF}$, for the case of the 
two-sample comparison of means ($t$-test) and the EQ scenario. Results for linear regression as well as for the UNEQ scenario 
were very similar and can be found in the online supplementary material. 

The performance of all \textit{Meta}-$BF_{g}$ methods except for the limited
information \textit{Meta-$BF_{g}^{(L)}$} is very satisfactory for the settings studied - see Fig. \ref{fig:1}.
 This suggests that the known variance approximation of the 
$T_k$ statistics is very adequate and knowledge of the 
individual {$ss_{k}$}
is not of cardinal importance for the successful synthesis of a
 small to moderate number of $BF$s. From Fig. \ref{fig:1} 
we can observe a relative discordance for large values of $BF$s
corresponding to larger effect sizes (i.e. when $\beta=0.3$, on
 the top right corner of the graphs). Nevertheless, this 
does not seem to be of major importance in terms of evidence classification. 

For the limited information scenario, the effect of the 
availability of only two-sided evidence is obvious for 
\textit{Meta-$BF_{g}^{(L)}$}; see Fig. \ref{fig:1}.
When $\beta=0$, the combined evidence systematically
understate evidence for $H_0$ as a result of the shortcomings
of synthesizing undirected $p$-values (or test statistics) as
discussed in Section \ref{sec:Limited_info}. For $K=2$ the
method performs surprisingly well for $\beta>0$. But as $K$
increases, it overstates the evidence for $H_1$ an effect partly mitigated for increasing $\beta$, as then, the number of $T_k$s with concordant signs naturally increases. 

\subsection{Random sample size}
\subsubsection{Simulation Plan}
For this simulation scenario, the total sample size was drawn
from a discrete $Uni(800,10000)$ and was split in $K$ parts of
random size, where $K \in \{2,\dots,50\}$. Model parameters were
the same as in the fixed sample size case. We
compared all \textit{Meta-$BF$} methods according to their
empirical bias, Root Mean Squared Error (RMSE) as well as the 
weighted Cohen's $\kappa$ statistic. For the 
calculation of the empirical bias and the RMSE, the $BF$ was treated as a 
parameter while the respective $\widetilde{BF}$ as the 
estimator. Therefore, for each simulation scenario:
\begin{align*}
Bias&=\frac{\sum_{j=1}^{R}(\widetilde{BF}-BF)}{R} \\
RMSE&=\sqrt{\frac{\sum_{j=1}^{R}(\widetilde{BF}-BF)^2}{R}}
\end{align*} 
where $R$ the number of simulations. The limited information estimator
\textit{Meta-$BF_{g}^{(L)}$} was not included in this 
comparison as it utilizes substantially less information 
from each study and, as it is shown in Section \ref{sec:res1} 
it does not perform satisfactorily for a
large number of studies. 

Comparisons are made on the $2\log BF$ scale. $\widetilde{BF}$s
 based on the $g$-prior and $JZS$ prior are compared to the 
 $g$-prior-based and $JZS$ prior-based $BF$ respectively.

\subsubsection{Results}
\begin{figure}
\includegraphics[height=0.95\textheight]{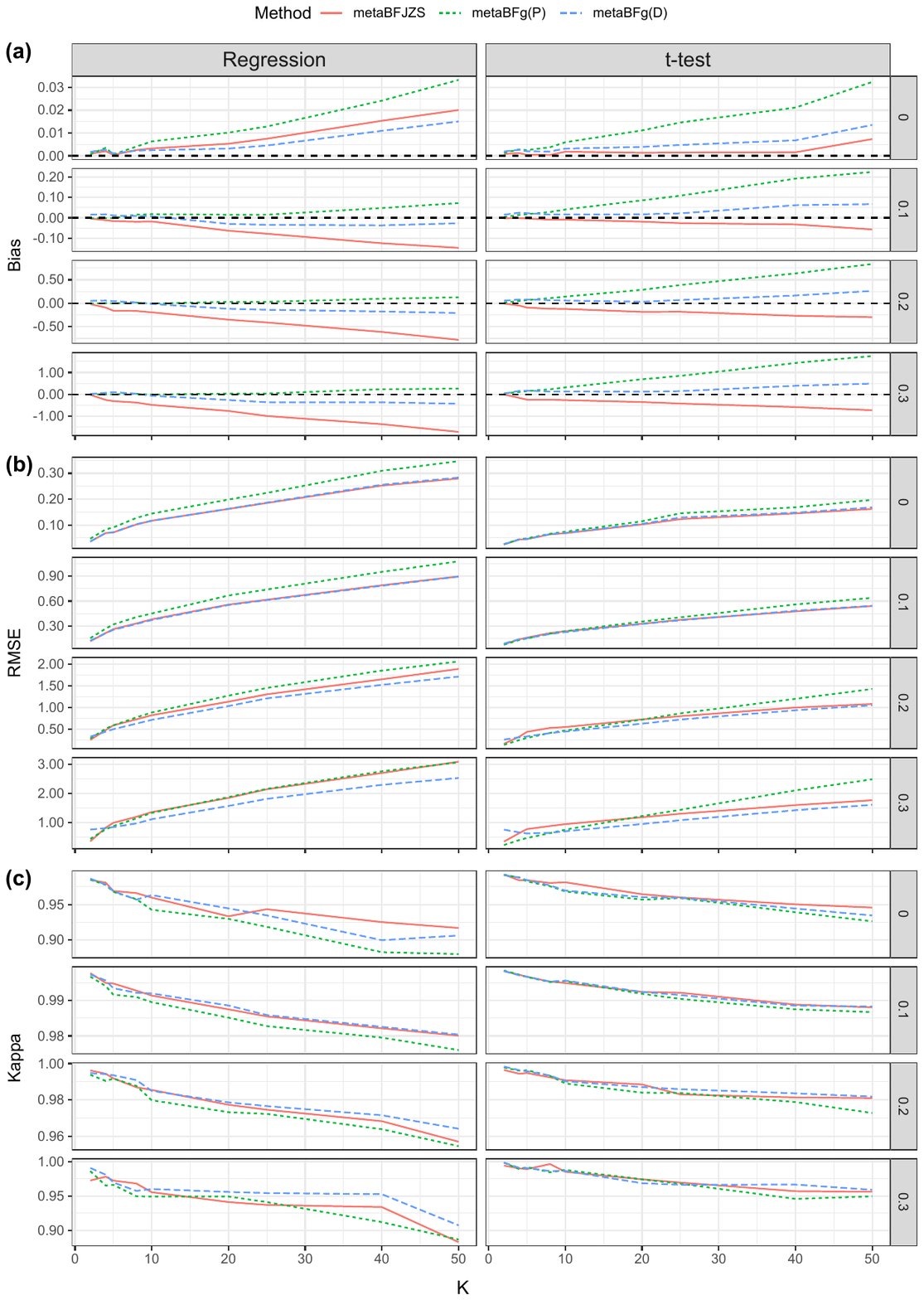}
\caption{Bias, RMSE and weighted Cohen's $\kappa$ versus $K$ (number
of subsamples) for random sample sizes. The scale is $2\log BF$. Columns indicate different models (regression and $t$-test) while sub-rows within each metric correspond to different values of $\beta$. The value of 0 for the Bias is highlighted by a dashed line.}
\label{fig:2}
\end{figure}

Figure \ref{fig:2} presents the results for the bias, RMSE and 
weighted Cohen's $\kappa$ statistic for increasing $K$, when the total sample size as well as $\omega_l$ for 
subsample $k\in \{1,...,K\}$ is random. For the two-sample comparison
of means ($t$-test), the effect of the lack of appropriate 
weighting imposed by  \textit{Meta-$BF_{g}^{(P)}$} becomes more 
apparent as $K$ grows. It exhibits higher bias	and RMSE.
Nevertheless, when it comes to the evidence quantification in a 
discrete manner, its performance is more than adequate, as 
indicated by the large ($>0.9$) values of the weighted Cohen's 
$\kappa$ statistic. The methods that utilize information about $ss_k$
, the coefficient's sun of squares in each subsample 
(\textit{Meta-$BF_{g}^{(D)}$} and \textit{Meta-$BF_{JZS}$}),
 are not substantially influenced by increasing $K$, with a 
notable difference on the bias: \textit{Meta-$BF_{JZS}$} 
exhibits negative bias, while \textit{Meta-$BF_{g}^{(D)}$} 
yields on average larger values than $BF$, with the bias being 
of slightly smaller absolute value compared to \textit{Meta-$BF_{JZS}$}.
\textit{Meta-$BF_{g}^{(D)}$} performs marginally better in 
terms of RMSE, for increasing $\beta$.

Results for the regression case are slightly different. One notable
difference is that $ss_k=(n_k-1)\hat{\sigma^2_{X_{1}}}$ and thus exhibiting
more variability than in the $t$-test case where $ss_k=\frac{n_k}{2}$.
This increased heterogeneity embodied in the $ss_k$, 
is handled in a better way by the \textit{Meta-$BF_{g}^{(D)}$} 
approach compared to \textit{Meta-$BF_{JZS}$}, in terms of bias
 and RMSE.  Such a behavior does not come as a surprise given
that the meta-analytic modeling is only partially implemented on the 
(variable) non-centrality parameter
of the individual $t$ statistics in the \textit{Meta-$BF_{JZS}$}
formulation.
 For the $t$-test, a comparable situation would be 
the case where the allocation ratios among the studies 
were different and not all equal to one as we have considered
in this simulation. Note that the \textit{Meta-$BF_{JZS}$} 
approach is performing better under $H_0$ in specific metrics 
(concerning the bias in the $t$-test and the Cohen's $\kappa$ in
regression). This may be attributed to the 
smaller variability in the (observed) non-centrality parameters 
imposed by $\beta=0$.

\section{Illustrative Example}\label{sec:Ex}
We apply the methods discussed in a meta-analysis from the field
of positive psychology. In \citet{bolier2013}, the authors present a 
systematic review and meta-analysis of randomized controlled 
trials assessing the effects of positive psychology 
interventions.
 The use of positive psychological interventions may be 
considered as a complementary strategy in mental health 
promotion and treatment. The efficacy of the interventions was 
assessed in several outcomes. Here, we focus on the outcome of 
psychological well being which exhibited low levels of 
heterogeneity and thus a common effect model is deemed appropriate. 

There were 20 studies included in the meta analysis. Since all measurements were continuous but 
instruments (i.e. questionnaires) varied across studies, 
the parameter of interest was the standardized mean difference. 
Table \ref{tab:3} shows the studies' results in terms of $T_k$ 
statistics, as well as the sample sizes per arm. We have added 
the calculated $BF_k$s using the $g$-prior ($BF_{k}^{(g)}$) as well 
as the ones using the $JZS$ prior $(BF_k^{(JZS)})$. The last two columns
 show the resulting (squared) weights of the studies, for 
the \textit{Meta-$BF_{g}^{(L)}$}/\textit{Meta-$BF_{g}^{(P)}$} as well as for the 
\textit{Meta-$BF_{g}^{(D)}$} approach.

Table \ref{tab:3} presents Bayes Factors in the form $BF_{10}$.
 The relationship between the $BF^{(g)}$s and 
$BF^{(JZS)}$ can be sketched. In general, $BF^{(JZS)} > BF^{(g)}$,
 except for when the sample size is relatively small, in which 
case the $BF^{(JZS)}$ is more conservative (e.g. Study 5 ). 
$BF^{(JZS)}$ gives less support for $H_0$ even when observed 
differences are very close to 0 (Studies 6 \& 19). 

For the $\widetilde{BF}_{g}$s, differences in the weighting 
schemes are also of interest. When the weights take into 
account the allocation ratio within the studies as well as 
the observed effects ($w_k^{(v)}$), both factors that 
contribute to the variance and mean of the $t$-statistics, 
studies with largely unequal allocation ratio are downweighted
compared to the weights based solely on the total sample sizes
of the studies ($w_k^{(\omega)}$) . Take for example studies 
12 \& 17: They are comparable with respect to their relative
total sample sizes (364 vs 411) but only 70/411 from study 
17 were allocated to group 2. This has considerable impact 
on its relative weight as $w_{17}^{(v)}$=0.097 while 
$w_{12}^{(v)}$=0.15.

\begin{table}[h!] 
	\centering
\def\arraystretch{1.5}	
		\begin{tabular}{rrrrrccccrr}
		\hline
		\hline
Study &	$T_k$ &	$n_{k(1)}$ &	$n_{k(2)}$ &	$n_k$ &$2\log BF^{\scriptscriptstyle(g)}$ &	$2 \log BF^{\scriptscriptstyle(JZS)}$ &	$BF^{\scriptscriptstyle(g)}$ &	$BF^{\scriptscriptstyle(JZS)}$ &	$w_k^{(\omega)}$ &	$w_k^{(v)}$ 	\\
\hline
1	  & -0.22	& 26	& 27	& 53	& -3.9	& -3.1	& 0.1	& 0.2	& 0.015	& 0.022	\\
2	  & 0.50  & 17	& 15	& 32	& -3.2	& -2.5	& 0.2	& 0.3	& 0.009	& 0.013	\\
3	  & 2.58	& 33	& 32	& 65	& 2.1	  & 2.5	  & 2.9	& 3.5	& 0.019	& 0.026	\\
4	  & 0.71	& 10	& 10	& 20	& -2.5	& -1.9	& 0.3	& 0.4	& 0.006	& 0.008	\\
5	  & 3.6	  & 8	  & 8	  & 16	& 6.2	  & 5.4	  & 22.2& 14.8& 0.005	& 0.004	\\
6	  & 0.00  & 37	& 29	& 66	& -4.2	& -3.3	& 0.1	& 0.2	& 0.019	& 0.027	\\
7	  & 0.82	& 79	& 86	& 165	& -4.4	& -3.6	& 0.1	& 0.2	& 0.047	& 0.069	\\
8	  & 0.77	& 559	& 63	& 622	& -5.8	& -3.9	& 0.1	& 0.1	& 0.178	& 0.095	\\
9	  & -0.08	& 35	& 38	& 73	& -4.3	& -3.4	& 0.1	& 0.2	& 0.021	& 0.030	\\
10	& 1.72	& 25	& 25	& 50	& -1.1	& -0.5	& 0.6	& 0.8	& 0.014	& 0.020	\\
11	& 0.86	& 80	& 37	& 117	& -4.0  & -3.1	& 0.1	& 0.2	& 0.033	& 0.042	\\
12	& 2.19	& 187	& 177	& 364	& -1.1	& -0.3	& 0.6	& 0.9	& 0.104	& 0.150	\\
13	& 2.68	& 153	& 89	& 242	& 1.6	  & 2.4	  & 2.2	& 3.3	& 0.069	& 0.093	\\
14	& 1.44	& 48	& 54	& 102	& -2.6	& -1.8	& 0.3	& 0.4	& 0.029	& 0.042	\\
15	& 1.21	& 804	& 138	& 942	& -5.4	& -3.8	& 0.1	& 0.1	& 0.269	& 0.198	\\
16	& 1.35	& 13	& 10	& 23	& -1.4	& -1.0  & 0.5	& 0.6	& 0.007	& 0.009	\\
17	& 0.11	& 341	& 70	& 411	& -6.0  & -4.5	& 0.0 & 0.1	& 0.117	& 0.097	\\
18	& 1.84	& 11	& 9	  & 20	&  0.1	& 0.2	  & 1.0 & 1.1	& 0.006	& 0.007	\\
19	& 0.00  & 36	& 42	& 78	& -4.4	& -3.5	& 0.1	& 0.2	& 0.022	& 0.032	\\
20	& 0.51	& 20	& 17	& 37	& -3.4	& -2.6	& 0.2	& 0.3	& 0.011	& 0.015	\\
\hline
\end{tabular}
\caption{Study result from \citet{bolier2013}, alongside the resulting $BF$s and weights based on sample fraction as well as on $v_k$}
\label{tab:3}	
\end{table}

The results for the meta-analytic $\widetilde{BF}$s are presented
in Table \ref{tab:4}. $\widetilde{BF}_{JZS}$ results
 to a considerably larger value, but belonging in the same 
evidence level as $\widetilde{BF}_{g}$ with respect to Table 
\ref{tab:1}, indicating very strong evidence for $H_1$.
The two methods which employ less data information 
result to slightly smaller $BF$s ($2\log BF \approx 9$).
 We see that \textit{Meta-$BF_{g}^{(L)}$} performs very well in
 this case. This is due to the fact that
 only two of the studies demostrated a $T$ statistic in the
 opposite direction. Based on the simulation results one would
 expect \textit{Meta-$BF_{g}^{(P)}$} to exhibit a larger value 
in comparison to \textit{Meta-$BF_{g}^{(D)}$}, but the effect of the discrepant weights due to combinations of unequal allocations and observed effect sizes results to the opposite.  

\begin{table}[h!]
	\centering
\def\arraystretch{1.5}	
		\begin{tabular}{ccc}
		\hline
		\hline
	Method & $\widetilde{2\log BF}$ & $\widetilde{BF}$ \\
	\hline
\textit{Meta-$BF_{g}^{(P)}$} &	9.00 &90.1\\
\textit{Meta-$BF_{g}^{(L)}$}&	9.64 &124.3\\
\textit{Meta-$BF_{g}^{(D)}$} &	11.83 &372.1\\
\textit{Meta-$BF_{JZS}$}  &	12.75 & 588.9 \\
\hline
\end{tabular}
\caption{Meta-analytic Bayes Factors for the illustrative 
example of Section \ref{sec:Ex}}
 \label{tab:4}	
\end{table}	

\section{Discussion}

In this work we explored methods for synthesizing aggregate level data in 
the form of Bayes Factors for the common effect model. We present methods
 for varying levels of detail in the available information and 
compare their performance. By utilizing the common $g$-prior 
setup, we show how the problem of $BF$ synthesis can be 
translated to the synthesis of $T$ statistics for simple 
linear regression. An analytic relationship between $BF$'s 
and standard test statistics can be found in several hypothesis
 testing contexts, outside the $g$-prior setup 
\citep{johnson2005,johnson2008,held2018}. 
This is also the case for the $JZS$ prior though it involves 
numerical integration. 

The $JZS$-prior can be seen as a mixture of $g$ priors. Alongside popularity, computational efficiency and understandable interpretation of $g$-priors, resulting $BF$s have been shown to be monotone functions of the $F$ statistic for testing \eqref{eq:2.1} and therefore gives the uniformly most powerful invariant test \citep{shively2018}. This holds also for the mixture of $g$-priors, and it is evident via the results presented here for the univariate regression example, considering that $F=T^2$. 

Meta-analysis is mostly involved with parameter estimates. The 
synthesis of $BF$s involves the synthesis of evidence concerning
 hypothesis testing. In frequentist inference, the analogous 
would be $p$-value combination methods. $p$-value 
combination methods are meant to result to a statistic that 
has a known distributional  form under the intersection of all 
null hypotheses being  tested $H_0=\cap_{i=1}^K\{H_{0i}\}$. 
Their purpose is to retain frequentist testing 
operational characteristics (type I error control) and not to 
synthesize test statistics. Therefore, a combined $p$-value should not be transformed to a 
meta-analytic Bayes Factor. We mitigate this shortcoming by employing 
the weighted test statistic from Stouffer's method, which, for 
proper weights (that sum to 1) and original $T_k$'s that are 
(approximately) normal, result to a correct 
$\widetilde{T}$ and $\widetilde{BF}$ for one-sided 
statistics (it is equivalent to the inverse normal method).

Our demonstration is limited to the case of a single covariate. Such a task could be very useful in simple but very common settings, as is the synthesis of evidence from a series of experiments, where testing is usually concerned with a difference between groups (e.g. randomized controlled trials). For more general problems where $BF$s are commonly employed, like variable/model selection, the extension is not straightforward. A similar strategy could be sought where regression slopes arising from independent datasets would be synthesized, though such a task does not come without complexity \citep{becker2007}. Such complexities include not reported covariance matrices as well as systematically missing covariates. 

We present meta-analytic Bayes Factors for varying levels of detail in the reported data from available studies. It is highly likely that
in realistic scenarios one can encounter a combination of such levels, i.e. some studies reporting only sample sizes and undirectional statistics, while others all the required ingredients for the computation of \textit{Meta-$BF_{g}^{(D)}$} or \textit{Meta-$BF_{JZS}$}. The reasonable choice in such cases would be to synthesize the evidence with a method that balances between using the most of the available information and is supported by some theoretical justification. For example, if information on the $ss_k$ is missing for some $k$ it is not straightforward how the weights should be differentiated between studies with and without available $ss_k$, and perhaps the preferred solution would be to proceed with weighting all studies by the sample sizes.

A limitation of our work is the treating only the common effect model, i.e. absence of heterogeneity. Heterogeneity is mostly defined as modeled variability on the true treatment effect level (ie the random-effects model), though it has also been referred to as a treatment $\times$ study interaction, ie the fixed effects model as defined here \citep{EMA2001MA}. Hypothesis testing for the random effects model involves several considerations which are further perplexed by the employment of Bayes Factors \citep{higgins2009}. The study of Meta-Anaytic Bayes Factors under the random-effects model requires considerable attention beyond the scope of this work. Our work is also focusing on point null hypotheses. Combining evidence for more complicated models is an active field of research \citep{kuiper_aggregating_2025, grunwald_beyond_2024,ramdas_game-theoretic_2023}.

The difficulties arising in the computation of $BF$s makes meta-analytic methods also relevant in data partitioning for Big Data when computations become CPU-intensive as well as data-intensive and cannot be handled by a single machine \citep{zhao2013}. $P$-value combination and other meta-analytic methods have been proposed in this field in order to tackle parallel processing of data partition for frequentist analyses \citep{tsamardinos2019}.  

Utilizing frequentist techniques in order to acquire Bayesian evidence may leave a bitter taste to Bayesians. A fully Bayesian approach would require placing a likelihood model on the $BF$s which is not straightforward. One could take advantage of results concerning the sampling distribution of $BF$s (a shifted non-central chi-square for $g$ priors, see \cite{zhou2018}) and pursue inference based on a full posterior or predictive distribution. However, such an exercise would strip $BF$s of the attractive and easily communicable feature of a single piece of evidence for or against a model.

Even though the incorporation of Bayes Factors in psychological research has been increasing \citep{heck2023review}, a reality where $BF$s are the only level of evidence available in published studies (even more $BF$s employing the same priors) is not close. However, the translation and interplay between frequentist and Bayesian evidence is of interest and meta-analyses have pointed towards re-assessing frequentist evidence through a Bayesian prism \citep{monden2016,bittl2019}. The methods developed and assessed in this work are fit towards this direction.

\printbibliography

@article{lehmacher1999,
  title={Adaptive sample size calculations in group sequential trials},
  author={Lehmacher, Walter and Wassmer, Gernot},
  journal={Biometrics},
  volume={55},
  number={4},
  pages={1286--1290},
  year={1999},
  publisher={Wiley Online Library}
}

@article{rice_re-evaluation_2018,
	title = {A {Re}-{Evaluation} of {Fixed} {Effect}(s) {Meta}-{Analysis}},
	volume = {181},
	copyright = {https://academic.oup.com/journals/pages/open\_access/funder\_policies/chorus/standard\_publication\_model},
	issn = {0964-1998, 1467-985X},
	url = {https://academic.oup.com/jrsssa/article/181/1/205/6884924},
	doi = {10.1111/rssa.12275},
	language = {en},
	number = {1},
	urldate = {2025-10-16},
	journal = {Journal of the Royal Statistical Society Series A: Statistics in Society},
	author = {Rice, Kenneth and Higgins, Julian P. T. and Lumley, Thomas},
	month = jan,
	year = {2018},
	pages = {205--227},
}

@article{hedges_fixed-_1998,
	title = {Fixed- and random-effects models in meta-analysis.},
	volume = {3},
	issn = {1939-1463, 1082-989X},
	url = {https://doi.apa.org/doi/10.1037/1082-989X.3.4.486},
	doi = {10.1037/1082-989X.3.4.486},
	language = {en},
	number = {4},
	urldate = {2025-10-16},
	journal = {Psychological Methods},
	author = {Hedges, Larry V. and Vevea, Jack L.},
	month = dec,
	year = {1998},
	pages = {486--504},
}

@article{carrera-rivera_how-conduct_2022,
	title = {How-to conduct a systematic literature review: {A} quick guide for computer science research},
	volume = {9},
	issn = {22150161},
	shorttitle = {How-to conduct a systematic literature review},
	url = {https://linkinghub.elsevier.com/retrieve/pii/S2215016122002746},
	doi = {10.1016/j.mex.2022.101895},
	urldate = {2025-10-09},
	journal = {MethodsX},
	author = {Carrera-Rivera, Angela and Ochoa, William and Larrinaga, Felix and Lasa, Ganix},
	year = {2022},
	pages = {101895},
}

@article{cucherat2000,
  title={Evidence of clinical efficacy of homeopathy},
  author={Cucherat, M and Haugh, MC and Gooch, M and Boissel, J-P},
  journal={European journal of clinical pharmacology},
  volume={56},
  number={1},
  pages={27--33},
  year={2000},
  publisher={Springer}
}

@article{higgins2009,
  title={A re-evaluation of random-effects meta-analysis},
  author={Higgins, Julian PT and Thompson, Simon G and Spiegelhalter, David J},
  journal={Journal of the Royal Statistical Society: Series A (Statistics in Society)},
  volume={172},
  number={1},
  pages={137--159},
  year={2009},
  publisher={Wiley Online Library}
}

@article{wang2017,
  title={Generalized R-squared for detecting dependence},
  author={Wang, Xufei and Jiang, Bo and Liu, Jun S},
  journal={Biometrika},
  volume={104},
  number={1},
  pages={129--139},
  year={2017},
  publisher={Oxford University Press}
}

@article{becker2007,
  title={The synthesis of regression slopes in meta-analysis},
  author={Becker, Betsy Jane and Wu, Meng-Jia},
  journal={Statistical Science},
  volume={22},
  number={3},
  pages={414--429},
  year={2007},
  publisher={Institute of Mathematical Statistics}
}

@article{engle1984,
  title={Wald, likelihood ratio, and Lagrange multiplier tests in econometrics},
  author={Engle, Robert F},
  journal={Handbook of Econometrics},
  volume={2},
  pages={775--826},
  year={1984},
  publisher={Elsevier}
}

@article{held2018,
  title={On p-values and Bayes factors},
  author={Held, Leonhard and Ott, Manuela},
  year={2018},
  journal={Annual Review of Statistics and Its Application},
	  volume={5},
		  pages={6.1-6.27},


}

@book{jeffreys1961,
  title={Theory of Probability (3rd edition)},
  author={Jeffreys, Harold},
  year={1961},
  publisher={Oxford University Press},
	address={}
}

@article{kass1995,
  title={Bayes factors},
  author={Kass, Robert E and Raftery, Adrian E},
  journal={Journal of the american statistical association},
  volume={90},
  number={430},
  pages={773--795},
  year={1995},
  publisher={Taylor \& Francis}
}

@article{liang2008,
  title={Mixtures of g priors for Bayesian variable selection},
  author={Liang, Feng and Paulo, Rui and Molina, German and Clyde, Merlise A and Berger, Jim O},
  journal={Journal of the American Statistical Association},
  volume={103},
  number={481},
  pages={410--423},
  year={2008},
  publisher={Taylor \& Francis}
}

@book{ohagan2004,
  title={Kendall's advanced theory of statistics, volume 2B: Bayesian inference},
  author={O'Hagan, Anthony and Forster, Jonathan J},
  volume={2},
  year={2004},
  publisher={Arnold},
	address={}
}

@article{owen2009,
  title={Karl Pearson’s meta-analysis revisited},
  author={Owen, Art B and others},
  journal={The Annals of Statistics},
  volume={37},
  number={6B},
  pages={3867--3892},
  year={2009},
  publisher={Institute of Mathematical Statistics}
}

@article{rouder2011,
  title={A Bayes factor meta-analysis of Bem’s ESP claim},
  author={Rouder, Jeffrey N and Morey, Richard D},
  journal={Psychonomic Bulletin \& Review},
  volume={18},
  number={4},
  pages={682--689},
  year={2011},
  publisher={Springer}
}

@article{shively2018,
  title={On Bayes factors for the linear model},
  author={Shively, T S and Walker, S G},
  journal={Biometrika},
  volume={105},
  number={3},
  pages= {739--744},
  year={2018},
  publisher={JSTOR}
}

@article{stouffer1949,
  title={The american soldier: Adjustment during army life.},
  author={Stouffer, Samuel A and Suchman, Edward A and DeVinney, Leland C and Star, Shirley A and Williams Jr, Robin M},
  year={1949},
  publisher={Princeton Univ. Press}
}

@article{tsamardinos2019,
  title={A greedy feature selection algorithm for Big Data of high dimensionality},
  author={Tsamardinos, Ioannis and Borboudakis, Giorgos and Katsogridakis, Pavlos and Pratikakis, Polyvios and Christophides, Vassilis},
  journal={Machine learning},
  volume={108},
  number={2},
  pages={149--202},
  year={2019},
  publisher={Springer}
}

@article{zellner1986,
  title={On assessing prior distributions and Bayesian regression analysis with g-prior distributions},
  author={Zellner, Arnold},
  journal={Bayesian inference and decision techniques},
  year={1986},
  publisher={Elsevier Science}
}

@article{zhao2013,
  title={Massively parallel feature selection: an approach based on variance preservation},
  author={Zhao, Zheng and Zhang, Ruiwen and Cox, James and Duling, David and Sarle, Warren},
  journal={Machine learning},
  volume={92},
  number={1},
  pages={195--220},
  year={2013},
  publisher={Springer}
}

@article{zhou2018,
  title={On the null distribution of Bayes factors in linear regression},
  author={Zhou, Quan and Guan, Yongtao},
  journal={Journal of the American Statistical Association},
  volume={113},
  number={523},
  pages={1362--1371},
  year={2018},
  publisher={Taylor \& Francis}
}

@article{johnson2005,
  title={Bayes factors based on test statistics},
  author={Johnson, Valen E},
  journal={Journal of the Royal Statistical Society: Series B (Statistical Methodology)},
  volume={67},
  number={5},
  pages={689--701},
  year={2005},
  publisher={Wiley Online Library}
}

@article{johnson2008,
  title={Properties of Bayes factors based on test statistics},
  author={Johnson, Valen E},
  journal={Scandinavian Journal of statistics},
  volume={35},
  number={2},
  pages={354--368},
  year={2008},
  publisher={Wiley Online Library}
}

@book	{fisher1934,
  title={Statistical methods for research workers. },
  author={Fisher, RA},
  publisher={Oliver and Boyd},
  year={1934},
	address={Edinburgh}
}

@article{pearson1934,
  title={On a New Method of Determining" Goodness of Fit"},
  author={Pearson, Karl},
  journal={Biometrika},
  volume={26},
  number={4},
  pages={425--442},
  year={1934},
  publisher={JSTOR}
}

@article{malzahn2000,
  title={Nonparametric estimation of heterogeneity variance for the standardised difference used in meta-analysis},
  author={Malzahn, Uwe and B{\"o}hning, Dankmar and Holling, Heinz},
  journal={Biometrika},
  volume={87},
  number={3},
  pages={619--632},
  year={2000},
  publisher={Oxford University Press}
}

@article{monden2016,
  title={Toward evidence-based medical statistics: a Bayesian analysis of double-blind placebo-controlled antidepressant trials in the treatment of anxiety disorders},
  author={Monden, Rei and de Vos, Stijn and Morey, Richard and Wagenmakers, Eric-Jan and de Jonge, Peter and Roest, Annelieke M},
  journal={International Journal of Methods in Psychiatric Research},
  volume={25},
  number={4},
  pages={299--308},
  year={2016},
  publisher={Wiley Online Library}
}

@article{wetzels2012,
  title={A default Bayesian hypothesis test for correlations and partial correlations},
  author={Wetzels, Ruud and Wagenmakers, Eric-Jan},
  journal={Psychonomic bulletin \& review},
  volume={19},
  number={6},
  pages={1057--1064},
  year={2012},
  publisher={Springer}
}

@article{bittl2019,
  title={Bayes factor meta-analysis of the mortality claim for peripheral paclitaxel-eluting devices},
  author={Bittl, John A and He, Yulei and Baber, Usman and Feldman, Robert L and von Mering, Gregory O and Kaul, Sanjay},
  journal={JACC: Cardiovascular Interventions},
  volume={12},
  number={24},
  pages={2528--2537},
  year={2019},
  publisher={Elsevier}
}

@article{gonen2005,
  title={The Bayesian two-sample t test},
  author={G{\"o}nen, Mithat and Johnson, Wesley O and Lu, Yonggang and Westfall, Peter H},
  journal={The American Statistician},
  volume={59},
  number={3},
  pages={252--257},
  year={2005},
  publisher={Taylor \& Francis}
}

@article{rouder2009,
  title={Bayesian t tests for accepting and rejecting the null hypothesis},
  author={Rouder, Jeffrey N and Speckman, Paul L and Sun, Dongchu and Morey, Richard D and Iverson, Geoffrey},
  journal={Psychonomic bulletin \& review},
  volume={16},
  number={2},
  pages={225--237},
  year={2009},
  publisher={Springer}
}

@article{gronau2019,
  title={Informed Bayesian t-tests},
  author={Gronau, Quentin F and Ly, Alexander and Wagenmakers, Eric-Jan},
  journal={The American Statistician},
  pages={1--14},
  year={2019},
  publisher={Taylor \& Francis}
}

@article{zellner1980,
  title={Posterior odds ratios for selected regression hypotheses},
  author={Zellner, Arnold and Siow, Aloysius},
  journal={Trabajos de estad{\'\i}stica y de investigaci{\'o}n operativa},
  volume={31},
  number={1},
  pages={585--603},
  year={1980},
  publisher={Springer}
}

@article{pateras2018,
  title={Data-generating models of dichotomous outcomes: heterogeneity in simulation studies for a random-effects meta-analysis},
  author={Pateras, Konstantinos and Nikolakopoulos, Stavros and Roes, Kit},
  journal={Statistics in medicine},
  volume={37},
  number={7},
  pages={1115--1124},
  year={2018},
  publisher={Wiley Online Library}
}

@article{fleiss1969,
  title={Large sample standard errors of kappa and weighted kappa.},
  author={Fleiss, Joseph L and Cohen, Jacob and Everitt, Brian S},
  journal={Psychological bulletin},
  volume={72},
  number={5},
  pages={323},
  year={1969},
  publisher={American Psychological Association}
}

@book{higgins2019,
  title={Cochrane handbook for systematic reviews of interventions},
  author={Higgins, Julian PT and Thomas, James and Chandler, Jacqueline and Cumpston, Miranda and Li, Tianjing and Page, Matthew J and Welch, Vivian A},
  year={2019},
  publisher={John Wiley \& Sons},
	address={}
}

@article{bolier2013,
  title={Positive psychology interventions: a meta-analysis of randomized controlled studies},
  author={Bolier, Linda and Haverman, Merel and Westerhof, Gerben J and Riper, Heleen and Smit, Filip and Bohlmeijer, Ernst},
  journal={BMC public health},
  volume={13},
  number={1},
  pages={119},
  year={2013},
  publisher={BioMed Central}
}

@misc{mulder_bayesian_2024,
	title = {Bayesian {Evidence} {Synthesis}: {Safely} and {Efficiently} {Combining} {Statistical} {Evidence} in {Meta}-{Analyses}},
	copyright = {https://creativecommons.org/licenses/by/4.0/legalcode},
	shorttitle = {Bayesian {Evidence} {Synthesis}},
	url = {https://osf.io/6c3kj},
	doi = {10.31234/osf.io/6c3kj},
	urldate = {2025-07-18},
	publisher = {Center for Open Science},
	author = {Mulder, Joris and Van Aert, Robbie Cornelis Maria},
	month = oct,
	year = {2024},
}

@article{mckenzie_brief_2024,
	title = {A brief note on the random-effects meta-analysis model and its relationship to other models},
	volume = {174},
	copyright = {https://www.elsevier.com/tdm/userlicense/1.0/},
	issn = {0895-4356},
	url = {https://linkinghub.elsevier.com/retrieve/pii/S0895435624002488},
	doi = {10.1016/j.jclinepi.2024.111492},
	language = {en},
	urldate = {2025-07-18},
	journal = {Journal of Clinical Epidemiology},
	author = {McKenzie, Joanne E. and Veroniki, Areti Angeliki},
	month = oct,
	year = {2024},
	note = {Publisher: Elsevier BV},
	pages = {111492},
}

@article{ramdas_game-theoretic_2023,
	title = {Game-{Theoretic} {Statistics} and {Safe} {Anytime}-{Valid} {Inference}},
	volume = {38},
	issn = {0883-4237},
	url = {https://projecteuclid.org/journals/statistical-science/volume-38/issue-4/Game-Theoretic-Statistics-and-Safe-Anytime-Valid-Inference/10.1214/23-STS894.full},
	doi = {10.1214/23-sts894},
	number = {4},
	urldate = {2025-07-18},
	journal = {Statistical Science},
	author = {Ramdas, Aaditya and Grünwald, Peter and Vovk, Vladimir and Shafer, Glenn},
	month = nov,
	year = {2023},
	note = {Publisher: Institute of Mathematical Statistics},
}

@article{grunwald_beyond_2024,
	title = {Beyond {Neyman}–{Pearson}: {E}-values enable hypothesis testing with a data-driven alpha},
	volume = {121},
	copyright = {https://creativecommons.org/licenses/by-nc-nd/4.0/},
	issn = {0027-8424, 1091-6490},
	shorttitle = {Beyond {Neyman}–{Pearson}},
	url = {https://pnas.org/doi/10.1073/pnas.2302098121},
	doi = {10.1073/pnas.2302098121},
	language = {en},
	number = {39},
	urldate = {2025-07-18},
	journal = {Proceedings of the National Academy of Sciences},
	author = {Grünwald, Peter D.},
	month = sep,
	year = {2024},
	note = {Publisher: Proceedings of the National Academy of Sciences},
}

@article{veroniki_brief_2024,
	title = {A brief note on the common (fixed)-effect meta-analysis model},
	volume = {169},
	copyright = {https://www.elsevier.com/tdm/userlicense/1.0/},
	issn = {0895-4356},
	url = {https://linkinghub.elsevier.com/retrieve/pii/S0895435624000362},
	doi = {10.1016/j.jclinepi.2024.111281},
	language = {en},
	urldate = {2025-07-18},
	journal = {Journal of Clinical Epidemiology},
	author = {Veroniki, Areti Angeliki and McKenzie, Joanne E.},
	month = may,
	year = {2024},
	note = {Publisher: Elsevier BV},
	pages = {111281},
}

@article{klugkist_bayesian_2023,
	title = {Bayesian evidence synthesis for informative hypotheses: {An} introduction.},
	copyright = {http://www.apa.org/pubs/journals/resources/open-access.aspx},
	issn = {1939-1463, 1082-989X},
	shorttitle = {Bayesian evidence synthesis for informative hypotheses},
	url = {https://doi.apa.org/doi/10.1037/met0000602},
	doi = {10.1037/met0000602},
	language = {en},
	urldate = {2025-07-18},
	journal = {Psychological Methods},
	author = {Klugkist, Irene and Volker, Thom Benjamin},
	month = sep,
	year = {2023},
	note = {Publisher: American Psychological Association (APA)},
}

@article{kuiper_aggregating_2025,
	title = {Aggregating evidence for a central theory from diverse studies using the generalized order-restricted information criterion.},
	issn = {1939-1463, 1082-989X},
	url = {https://doi.apa.org/doi/10.1037/met0000755},
	doi = {10.1037/met0000755},
	language = {en},
	urldate = {2025-07-18},
	journal = {Psychological Methods},
	author = {Kuiper, R. M. and Clapper, Eli-Boaz},
	month = may,
	year = {2025},
	note = {Publisher: American Psychological Association (APA)},
}

@misc{zhang_towards_2023,
	title = {Towards more scientific meta-analyses},
	copyright = {Creative Commons Attribution 4.0 International},
	url = {https://arxiv.org/abs/2308.13514},
	doi = {10.48550/ARXIV.2308.13514},
	urldate = {2025-07-18},
	publisher = {arXiv},
	author = {Zhang, Lily H. and Konstantinidis, Menelaos and Bind, Marie-Abèle and Rubin, Donald B.},
	year = {2023},
	note = {Version Number: 1},
}

@article{van_wonderen_bayesian_2024,
	title = {Bayesian evidence synthesis as a flexible alternative to meta-analysis: {A} simulation study and empirical demonstration},
	volume = {56},
	copyright = {https://creativecommons.org/licenses/by/4.0},
	issn = {1554-3528},
	shorttitle = {Bayesian evidence synthesis as a flexible alternative to meta-analysis},
	url = {https://link.springer.com/10.3758/s13428-024-02350-2},
	doi = {10.3758/s13428-024-02350-2},
	language = {en},
	number = {4},
	urldate = {2025-07-18},
	journal = {Behavior Research Methods},
	author = {Van Wonderen, Elise and Zondervan-Zwijnenburg, Mariëlle and Klugkist, Irene},
	month = mar,
	year = {2024},
	note = {Publisher: Springer Science and Business Media LLC},
	pages = {4085--4102},
}

@article{van_lissa_tutorial_2024,
	title = {A tutorial on aggregating evidence from conceptual replication studies using the product {Bayes} factor},
	volume = {15},
	copyright = {http://creativecommons.org/licenses/by/4.0/},
	issn = {1759-2879, 1759-2887},
	url = {https://onlinelibrary.wiley.com/doi/10.1002/jrsm.1765},
	doi = {10.1002/jrsm.1765},
	language = {en},
	number = {6},
	urldate = {2025-07-18},
	journal = {Research Synthesis Methods},
	author = {Van Lissa, Caspar J. and Clapper, Eli‐Boaz and Kuiper, Rebecca},
	month = nov,
	year = {2024},
	note = {Publisher: Wiley},
	pages = {1231--1243},
}

@article{heck2023review,
  title={A review of applications of the Bayes factor in psychological research.},
  author={Heck, Daniel W and Boehm, Udo and B{\"o}ing-Messing, Florian and B{\"u}rkner, Paul-Christian and Derks, Koen and Dienes, Zoltan and Fu, Qianrao and Gu, Xin and Karimova, Diana and Kiers, Henk AL and others},
  journal={Psychological Methods},
  volume={28},
  number={3},
  pages={558},
  year={2023},
  publisher={American Psychological Association}
}

@misc{EMA2001MA,
	author       = {{European Medicines Agency}},
	title        = {Points to consider on application with 1. meta-analyses; 2. one pivotal study},
	howpublished = {Guideline, CPMP/EWP/2330/99},
	institution  = {European Medicines Agency (EMA), London},
	year         = {2001},
	month        = may,
	day          = {31},
	note         = {Adopted by the Committee for Proprietary Medicinal Products (CPMP)},
	url          = {https://www.ema.europa.eu/en/documents/scientific-guideline/points-consider-application-1meta-analyses-2one-pivotal-study_en.pdf}
}
\end{document}